\pdfoutput=1
\documentclass[11pt]{article}
\usepackage[final]{acl}
\usepackage{times}
\usepackage{latexsym}
\usepackage[T1]{fontenc}
\usepackage[utf8]{inputenc}
\usepackage{microtype}
\usepackage{inconsolata}
\usepackage{amsmath,amsthm,amssymb,graphicx}
\usepackage{multirow} 
\usepackage{algorithm}
\usepackage{algorithmic}
\usepackage{setspace}
\usepackage{booktabs}
\usepackage{pgfplots}
\usepackage{subfigure}
\usepackage{pifont}
\usepackage{CJKutf8} 
\usepackage{tabularx}
\usepackage{array}

\title{VCB Bench: An Evaluation Benchmark for Audio-Grounded Large Language Model Conversational Agents}

\author{
  Jiliang Hu\textsuperscript{1,2,\footnotemark[1]},
  Wenfu Wang\textsuperscript{2,\footnotemark[2]},
  Zuchao Li\textsuperscript{1,\footnotemark[2],\footnotemark[3]},
  Chenxing Li\textsuperscript{2},
  Yiyang Zhao\textsuperscript{2}
  \\
  \bfseries
  Hanzhao Li\textsuperscript{2},
  Liqiang Zhang\textsuperscript{2},
  Meng Yu\textsuperscript{2},
  Dong Yu\textsuperscript{2}
  \\
  \textsuperscript{1}School of Artificial Intelligence, Wuhan University, Wuhan, China, \\
  \textsuperscript{2}Tencent AI Lab, Beijing, China. \\
  \{jilianghu, zcli-charlie\}@whu.edu.cn, \\
  \{wenfuwang, chenxingli, yyangyzhao, hanzhaoli, tatelqzhang\}@tencent.com, \\
  \{raymondmyu, dyu\}@global.tencent.com.
}

\begin{document}
\maketitle
\begin{abstract}

While large audio language models (LALMs) have driven significant progress in multimodal conversational systems, current benchmarks suffer from critical limitations: they are largely English-centric, use synthetic speech, and fail to provide comprehensive, discriminative evaluation across key dimensions. To fill this gap, we present Voice Chat Bot Bench (VCB Bench), a novel, high-quality Chinese benchmark built exclusively on real human speech. VCB Bench assesses LALMs across three complementary axes: instruction following (including speech-level control beyond text commands), knowledge understanding (including general knowledge, reasoning, and daily dialogue), and robustness (evaluating stability under variations in content, environment, and speaker characteristics). Experiments conducted on representative LALMs reveal notable performance disparities and offer tangible insights for future improvements. VCB Bench serves as a reproducible and fine-grained framework, providing standardized evaluation and practical guidance for the development of Chinese voice conversational models. \footnote{Code and data are available at \url{https://github.com/Tencent/VCB-Bench}}
\end{abstract}

\renewcommand{\thefootnote}{\fnsymbol{footnote}}
\footnotetext[1]{This work was done when the first author worked as an intern at Tencent AI Lab.}
\footnotetext[2]{Corresponding authors.}
\footnotetext[3]{This work was supported by the National Natural Science Foundation of China (No. 62306216).}
\renewcommand {\thefootnote}{\arabic{footnote}}


\section{Introduction}

In recent years, large language models (LLMs) \citep{vaswani2017attention,anil2023palm} have achieved remarkable progress in natural language understanding and generation. Integrating language modeling with modalities such as vision and audio \citep{radford2021learning,singh2022flava}  has further given rise to a new paradigm of multimodal learning. Within this trend, LALMs---which combine speech signal processing with language modeling---have developed rapidly. Emerging systems such as Step-Audio 2 \citep{wu2025step2} and Qwen3-Omni \citep{xu2025qwen3} demonstrate end-to-end (E2E) speech understanding and generation with capabilities in voice question answering, real-time conversation, and audio content analysis. Consequently, voice conversational agents powered by LALMs are drawing increasing academic and industrial attention, offering more natural and human-like interactions than text-only systems.


Despite these advances, the transition from basic LALM functionalities to practical voice agents necessitates reliable and comprehensive evaluation tools. Such benchmarks are essential for diagnosing model weaknesses, guiding optimization efforts, and enabling fair comparisons across different systems. While initial efforts \citep{chen2024voicebench,yang2024air,lin2025full} have explored instruction following, audio understanding, reasoning, and dialogue scenarios, current evaluation practices remain limited in three major aspects: First, existing benchmarks are predominantly English-centric, resulting in insufficient exploration of Chinese, despite it being the world’s most widely spoken language. Second, the majority rely on synthetic speech data, which poorly reflects the acoustic variability of real-world environments. Third, many are derived from text-based benchmarks (e.g., AlpacaEval \citep{li2023alpacaeval}, IFEval \citep{zhou2023instruction}). The formal and lengthy content of these benchmarks is unsuitable for evaluating conversationally grounded LALMs, which are designed to generate natural, colloquial speech. This effort is critical due to China’s vast user base and the rapidly growing demand for practical, high-quality voice agents.


We address these limitations by introducing VCB Bench: the first comprehensive evaluation framework for Chinese voice conversation, built entirely upon authentic (non-synthetic) speech. VCB Bench assesses LALMs across three complementary dimensions: (1) Instruction Following: This extends beyond text-based prompts to include speech-level control tasks, such as adjusting volume, speed, and emotion, with inherent bilingual (Chinese-English) support. (2) Knowledge: This encompasses multi-disciplinary general knowledge (across 12 subjects), mathematical and logical reasoning, daily dialogue comprehension, and story continuation to assess pre-training performance. (3) Robustness: This measures the model’s stability under real-world perturbations, including variations in content (mispronunciations, grammatical errors), environment (street, TV noise), and speaker characteristics (age, accents).






Our proposed VCB Bench is entirely derived from authentic human recordings, moving beyond the limitations of synthetic speech. It provides a large-scale, high-fidelity dataset encompassing diverse conversational scenarios and introduces a multi-dimensional evaluation framework. This framework jointly measures instruction following (including fine-grained control), knowledge understanding, and robustness through a set of reproducible tasks. Based on this benchmark, we conduct a systematic empirical analysis of state-of-the-art LALMs under unified settings, which elucidates their strengths and limitations in Chinese voice interaction and yields actionable insights for future model development.

\section{Related Work}


\noindent \textbf{Large Audio Language Models.}\hspace{0.5em} LALMs primarily adopt an E2E audio-language modeling paradigm \citep{tang2023salmonn,fu2024vita,xie2024mini}, integrating speech understanding and generation within a unified framework. 

Qwen-Audio and Qwen-Omni series \citep{chu2023qwen,chu2024qwen2,xu2025qwen2,xu2025qwen3} progressively enhance cross-modal alignment and modeling efficiency. Qwen-Audio establishes robust audio-text alignment, Qwen-Audio2 improves encoding efficiency via multi-scale feature fusion, and the latest Qwen-Omni models introduce dual-core Thinker-Talker architectures and multi-codebook pretraining, achieving low-latency bilingual dialogue. 

StepAudio models \citep{huang2025step,wu2025step2} focus on tightly coupling recognition and synthesis. StepAudio integrates a dual-codebook tokenizer and achieves a remarkably low WER, while Step-Audio 2 mini advances to a fully E2E design with fixed text-speech token alignment and Chain-of-Thought reasoning, improving fine-grained paralinguistic understanding. 

\begin{figure*}[htbp]
	\centering 
	\includegraphics[scale=0.34]{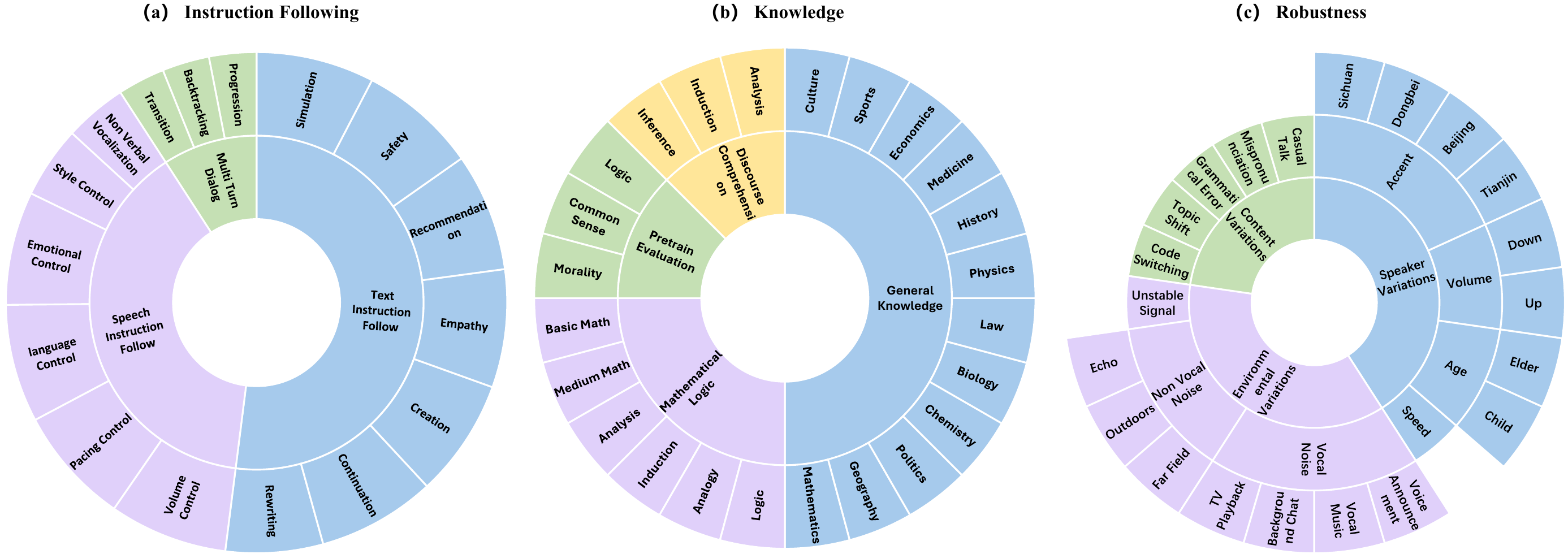} 
	\caption{Overview of VCB Bench.} 
    \label{distribution}
\end{figure*}

Baichuan-Audio \citep{li2025baichuan} employs hierarchical RVQ discretization and dual audio heads to balance acoustic and linguistic objectives, enabling real-time bilingual communication. GLM-4-Voice \citep{zeng2024glm} introduces a three-module structure (Tokenizer-Backbone-Decoder) supporting emotion and dialect modeling. Kimi-Audio \citep{ding2025kimi} fuses continuous acoustic and discrete semantic tokens in a dual-head architecture, achieving low-latency, high-fidelity streaming generation.


In addition, there are many other excellent LALMs, such as MiMo-Audio \citep{zhang2025mimo} and Fun-Audio-Chat \citep{chen2025fun}. All of these models demonstrate rapid progress in unified audio-language modeling---covering tokenization, multimodal fusion, and real-time dialogue---but systematic benchmarks, especially for Chinese real-speech interaction, remain scarce. Current evaluations are mostly qualitative or based on synthetic data, underscoring the need for comprehensive real-speech benchmarks like VCB Bench.

\noindent \textbf{Audio Benchmarks.}\hspace{0.5em} Recent efforts have introduced several benchmarks to evaluate LALMs from different perspectives. VoiceBench \citep{chen2024voicebench} assesses general knowledge, instruction adherence, safety, and robustness, primarily by adapting content from existing text datasets such as AlpacaEval and SD-QA \citep{faisal2021sd}, often after removing lengthy or overly complex samples. OpenAudioBench, released with Baichuan-Audio, integrates question-answering datasets like Spoken LLaMA Questions \citep{nachmani2023spoken} and Web Questions \citep{berant2013semantic}, and further augments them with a TTS-generated reasoning subset.

AIR Bench \citep{yang2024air} comprises two components: a basic benchmark covering emotion recognition, automatic speech recognition (ASR), and age estimation, and a dialogue benchmark focused on evaluating auditory understanding and internal knowledge. MMAU \citep{kumar2025mmau} and MMAR \citep{ma2025mmar} concentrate on deep audio reasoning, specifically requiring multi-step inference grounded in the model's internal audio knowledge. OmniBench \citep{li2024omnibench} targets omni-modal models that process audio, images, and text. It utilizes text queries paired with multimodal contexts (speech, music, or sound) to test integrated reasoning capabilities. Finally, URO Bench \citep{yan2025uro} provides a comprehensive bilingual (English-Chinese) set for evaluating audio understanding, reasoning, and conversational ability, but its speech data are entirely synthetic (TTS-generated).

Overall, existing benchmarks have significantly advanced the evaluation coverage of LALMs, yet they share several limitations:
(1) most rely heavily on TTS or synthetic speech, (2) they focus on English, and (3) their content often derives from text-centric QA corpora rather than spontaneous human dialogue. These gaps highlight the need for a real-speech, Chinese-oriented benchmark offering multi-dimensional evaluation---the central goal of our proposed VCB Bench. 


\section{VCB Bench}

As shown in the Figure \ref{distribution}, VCB Bench covers three core dimensions: Instruction Following, Knowledge, and Robustness. Instruction Following includes Text Instruction Following (TIF) (e.g., continuation, creation), Speech Instruction Following (SIF) (e.g., emotional, volume control), and Multi-turn Dialog (MTD) tasks. The Knowledge module assesses General Knowledge (GK) across 12 disciplines, Mathematical Logic (ML), Discourse Comprehension (DC), and Story Continuation (SC). Robustness introduces real-world perturbations from speaker variations, environmental noise, and content modifications to evaluate model stability. The detailed statistics for each subtask is presented in Table \ref{sumup}.

\subsection{Dataset Construction}

The VCB Bench dataset integrates data from three distinct sources: third-party professional recordings, audio segments extracted from variety show Q\&A, and an internally curated two-person conversational dialogue dataset. Each source is utilized to support different evaluation modules within the benchmark. The sources of variety show Q\&A data and the internal two-person dialogue dataset are detailed in Appendix \ref{src_of_data}.

In addition, during the dataset construction process, many procedures involved human annotators. For these procedures, we presented instructional guidelines to the participants. An example of such instructions is provided in Appendix \ref{data_ins}.



\begin{figure*}[htbp]
	\centering 
	\includegraphics[scale=0.54]{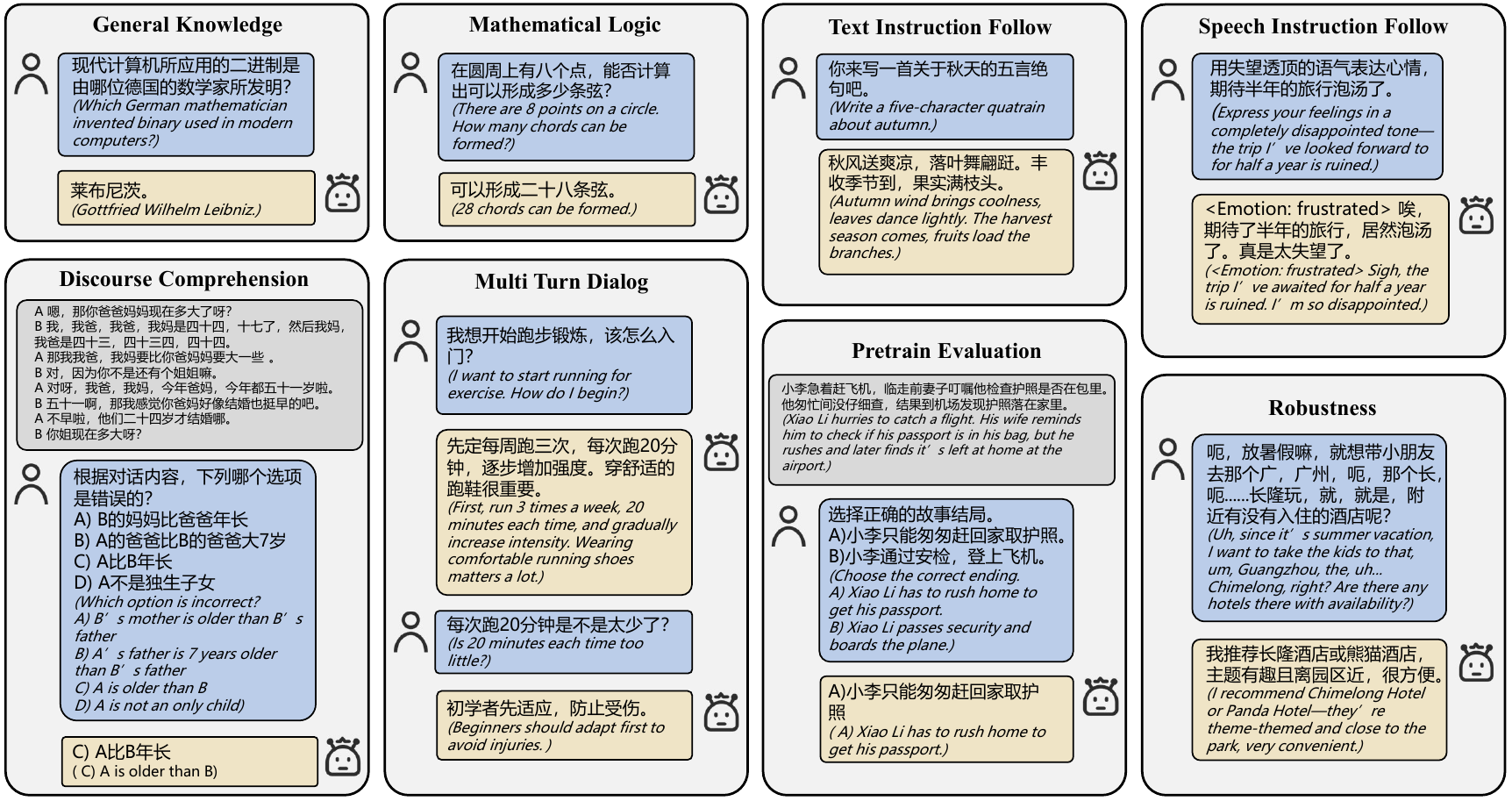}
	\caption{Examples from the VCB Bench.} 
    \label{examples}
\end{figure*}

\noindent \textbf{Third-Party Recorded Data.}\hspace{0.5em} This data category supports the Instruction Following, ML, and SC tasks in the Knowledge module, as well as the Robustness module.The production pipeline has several steps: First, task types and specific examples are defined. Next, professional writers manually craft task-compliant text. This text undergoes rigorous manual quality checks before being sent to a third-party team for professional audio recording. After recording, the data team first checks audio quality to filter out low-quality files. They then conduct detailed manual screening, selecting top-quality samples to form the final evaluation dataset.



The text materials for the Robustness data are derived directly from the Instruction Following module, with the original audio from that module serving as the control group. To control for speaker variability, the same speaker re-recorded the text under specified interference conditions (e.g., accent, noisy environment) wherever possible, using the original audio as a baseline. For content variation types, the text was first modified (e.g., introducing grammatical or pronunciation errors) before being re-recorded by the same speaker. Additionally, to test performance in extreme scenarios, subsets such as Volume, Speed, and Unstable Signal were created using post-processing techniques.



\noindent \textbf{Variety Show Q\&A Data.}\hspace{0.5em} This category supports GK in the Knowledge module. The process involves: crawling about 70 hours of Q\&A audio; manual timestamp annotation and segmentation; ASR transcription; question subject classification; and a final manual review for transcription accuracy, audio clarity, problem statement and answer correctness.

\noindent \textbf{Internal Two-person Dialogue Dataset.}\hspace{0.5em} Designed to support the DC module, this category's data is processed as follows: the original long-form audio undergoes a two-stage segmentation ---first by topic, then refined into semantically coherent segments under one minute. From the ASR transcriptions corresponding to the segments, we ask LLM to generate task-specific QA pairs (e.g., for analysis or induction), followed by a final manual screening to verify question quality and answer accuracy.

\subsection{Dataset Details}

\noindent \textbf{Instruction Following.}\hspace{0.5em} The Instruction Following section provides a comprehensive evaluation of LALMs' capacity to understand and execute both text and speech commands, encompassing three sub-tasks: TIF, SIF, and MTD. All tasks are open-ended, and both TIF and SIF support bilingual evaluation (Chinese and English) to address cross-lingual assessment requirements.

\begin{table*}[htbp]
\small
\begin{tabularx}{\textwidth}{lccX}
    \toprule
    \textbf{Dimension} & \textbf{Type} & \textbf{Size} & \textbf{Subsets (Size)} \\
    \midrule
    \multirow{5}{*}{Instruction Following}   & TIF & 1365 & Continuation (200), Creation (200), Empathy (200), Recommendation (200), Rewriting (165), Safety (200), Simulation (200) \\
    \cmidrule(lr){2-4}
    & TIF-En & 1354 & Continuation En (200), Creation En (200), Empathy En (200), Recommendation En (200), Rewriting En (154), Safety En (200), Simulation En (200) \\
    \cmidrule(lr){2-4}
    & SIF & 1020 &  Emotional Control (192), Language Control (200), Non Verbal Vocalization (106), Pacing Control (200), Style Control (122), Volume Control (200) \\
    \cmidrule(lr){2-4}
    & SIF-En & 1001 & Emotional Control En (173), Language Control En (200), Non Verbal Vocalization En (125), Pacing Control En (200), Style Control En (103), Volume Control En (200) \\
    \cmidrule(lr){2-4}
    & MTD & 240 & Progression (80), Backtracking (80), Transition (80) \\
    \midrule
    \multirow{4}{*}{Knowledge} & GK & 1041 & Mathematics (36), Geography (150), Politics (59), Chemistry (46), Biology (125), Law (37), Physics (102), History (150), Medicine (77), Economics (48), Sports (61), Culture (150) \\
    \cmidrule(lr){2-4}
    & ML & 663 & Basic Math (146), Medium Math (170), Analysis (84), Induction (64), Analogy (40), Logic (159)  \\
    \cmidrule(lr){2-4}
    & DC & 331 & Inference (103), Induction (113), Analysis (115) \\
    \cmidrule(lr){2-4}
    & SC & 379 & Logic And Causality (261), Common Sense And Science (72), Morality And Emotion (46) \\
    \midrule
    \multirow{3}{*}{Robustness} & SV & 349 & Age (95), Accent (108), Volume (100), Speed (46) \\
    \cmidrule(lr){2-4}
    & EV & 522 & Non Vocal Noise (150), Vocal Noise (292), Unstable Signal (80) \\
    \cmidrule(lr){2-4}
    & CV & 544 & Casual Talk (203), Mispronunciation (89), Grammatical Error (69), Topic Shift (91), Code Switching (92) \\
    \bottomrule
\end{tabularx}
\centering
\caption{Detailed statistics for each subtask on VCB Bench.}
\label{sumup}
\end{table*}

TIF assesses the model’s ability to respond to textual instructions through seven sub-tasks, each examining text generation and semantic comprehension: (1) Continuation: extending a given text fragment to evaluate coherence and creativity; (2) Creation: generating original content based on a given theme to assess inventiveness and organization; (3) Empathy: understanding and responding to emotional expressions to examine affective perception; (4) Recommendation: providing suggestions based on user needs to evaluate information integration; (5) Rewriting: adapting text in style or structure to test reorganization ability; (6) Safety: identifying and rejecting harmful instructions to assess compliant response; (7) Simulation: role-playing in dialogue to examine contextual adaptation.

SIF focuses on understanding and executing speech instructions, particularly the ability to handle paralinguistic features such as emotion, speaking rate, and dialect. It includes six sub-tasks: (1) Emotional Control: adjusting the emotional tone of speech to assess expressive generation; (2) Language Control: switching languages or dialects to test multilingual synthesis; (3) Non-verbal Vocalization: incorporating non-linguistic elements like sighs or nasal sounds to evaluate paralinguistic expressiveness; (4) Pacing Control: modifying speaking rate to examine control precision; (5) Style Control: switching speech styles to assess style transfer; (6) Volume Control: adjusting loudness to test stability.

MTD evaluates instruction tracking and topic management in multi-turn dialogues, each containing 3-5 turns, focusing on contextual understanding and logical coherence: (1) Progression: deepening the discussion around an initial topic to assess topic development; (2) Backtracking: recalling and responding to previously mentioned information to test long-range memory; (3) Transition: suddenly shifting to a new topic to evaluate conversational flow and relevance.

\noindent \textbf{Knowledge.}\hspace{0.5em} The Knowledge module evaluates LALMs’ knowledge storage, logical reasoning, and spoken dialog comprehension through reference-based question answering. It comprises four sub-tasks: GK, ML, DC, and SC.

GK evaluates multi-disciplinary common sense across twelve core domains---mathematics, geography, politics, chemistry, biology, law, physics, history, medicine, economics, sports, and culture---to measure the model's ability to recall and apply knowledge across diverse fields.

ML module consists of two key components: Mathematics and Logical Reasoning. Mathematics is divided into Basic Math, which is confined to integer arithmetic within 100, and Medium Math, which includes advanced algebra, geometry, number theory, and related disciplines, collectively assessing computational and problem-solving skills. Logical Reasoning comprises four reasoning types: Analysis for breaking down information, Induction for identifying and generalizing patterns, Analogy for mapping relational correspondences, and Logic for executing conditional reasoning, thereby testing analytical and deductive capabilities.

DC focuses on understanding dialogues through three dedicated tasks: Analysis detects factual accuracy within dialogues, Induction summarizes overarching dialogue themes, and Inference deduces speakers' attitudes, emotions, or intents, together evaluating comprehension and implicit reasoning skills.

SC, inspired by StoryCloze \citep{mostafazadeh2016corpus}, assesses implicit reasoning by requiring the model to select the correct story ending from two candidates, where both the context and the candidate endings are provided in the same modality---either all in audio or all in text. This task spans three evaluative categories: Logic and Causality for causal consistency, Common Sense and Science for real-world and scientific knowledge alignment, and Morality and Emotion for moral and emotional coherence.

\noindent \textbf{Robustness.}\hspace{0.5em} The Robustness module evaluates the stability of LALMs' performance under real-world interference conditions, ensuring reliable responses in challenging scenarios. The module encompasses three dimensions: Speaker Variations (SV), Environmental Variations (EV), and Content Variations (CV).

SV examine model adaptation to speaker attributes: (1) Age: utilizes child and elderly speech to assess recognition of age-related vocal characteristics; (2) Accent: incorporates four regional accents (Tianjin, Beijing, Dongbei, Sichuan) to evaluate comprehension of non-standard Mandarin; (3) Volume: assesses perception stability with amplified/attenuated speech; (4) Speed: tests parsing capability with rapidly delivered input.

EV simulate acoustic interference: (1) Non-Vocal Noise: includes echo, outdoor, and far-field noise; (2) Vocal Noise: contains television audio, background conversations, vocal music, and radio broadcasts; (3) Unstable Signal: emulates network-induced packet loss to evaluate handling of fragmented audio.


CV introduce linguistic disruptions: (1) Casual Talk: incorporate discourse markers (e.g., "um", "ah") and repeated phrases/words; (2) Mispronunciation: introduce phonetic deviations; (3) Grammatical Error: employ ungrammatical constructions; (4) Topic Shift: implement abrupt topic changes; (5) Code-Switching: mix Chinese and English. Each category evaluates the model's ability to maintain comprehension despite content imperfections.

\section{Experiment}
\subsection{Configuration}

We evaluate the latest and most capable LALMs. The selected models comprise GLM-4-Voice, Kimi-Audio, Qwen2.5-Omni, Baichuan-Audio, Step-Audio 2 mini, MiMo-Audio, GPT-4o-Audio, Qwen3-Omni and Fun-Audio-Chat.


For all tasks except SC, we generate spoken responses by calling each model's audio-to-audio API. For SIF tasks in both Chinese and English, the audio responses are evaluated directly using Gemini-2.5-Pro. For the remaining tasks, following prior work \citep{yan2025uro}, we first transcribe the audio responses using an ASR model (Whisper \citep{radford2023robust} for English tasks and Paraformer \citep{gao2022paraformer} for Chinese tasks) and then assess the transcriptions with GPT-4o. In open-ended QA, both Gemini and GPT assign a numerical score on a 1–5 scale, whereas for reference-based QA, they output a binary judgment of "Yes" or "No."

For the SC task, we assess a subset of pre-trained base models: Baichuan-Audio-Base, Kimi-Audio-Base, Qwen2-Audio-Base, and Step-Audio 2 mini-Base. Following the StoryCloze evaluation protocol, we compute the negative log-likelihood for both the correct and incorrect endings, with model selection determined by comparing these two values. For SIF tasks, six performing models undergo further Mean Opinion Score (MOS) evaluation. We sample the first 30 items from each relevant dataset, and eight expert evaluators rate the generated audio samples. 



\begin{table*}[htbp]
    \renewcommand{\arraystretch}{1.2} 
    \small
    \resizebox{\linewidth}{!}{
    \begin{tabular}{lcccccccccccc}
    \hline
    \multirow{2}{*}{Model}  &\multicolumn{5}{c}{Instruction Following} &\multicolumn{3}{c}{Knowledge} &\multicolumn{3}{c}{Robustness} \\
    \cmidrule(lr){2-6}\cmidrule(lr){7-9}\cmidrule(lr){10-12}
    & TIF & TIF-En & SIF & SIF-En & MTD & GK & ML & DC & SV & EV & CV\\
    \hline
    GLM-4-Voice \citep{zeng2024glm} & 82.15 & 75.52 & 73.18 & 66.94 & 82.56 & 41.79 & 60.18 & / & 73.64&77.51&78.60\\
    Kimi-Audio \citep{ding2025kimi} & 77.33 & 60.37 & 71.04 & 54.13 & 82.27 & 54.47 & 75.42 & 51.96&64.35&65.14&71.51 \\
    Qwen2.5-Omni \citep{xu2025qwen2}& 76.75 & 60.83 & 54.04 & 42.18 & 85.23 & 48.13 & 73.00 & 72.81&77.82&74.41&78.27 \\
    Baichuan-Audio \citep{li2025baichuan} & 82.24 & 78.60 & 53.67 & 48.27 & 80.96 & 40.73 & 74.05 & 51.66&78.05&77.62&78.71 \\
    Step-Audio 2 mini \citep{wu2025step2} &76.10& 70.47& 60.80& 49.85& 82.31& 49.86& 72.85& 79.46& 75.19& 70.58& 74.78\\
    MiMo-Audio \citep{zhang2025mimo} &90.08&81.89&56.26&42.74&86.30&48.70&81.75&/&83.72&85.36&\textbf{89.38} \\
    GPT-4o-Audio \citep{gpt4o_2024}&86.94&88.80&77.98&\textbf{82.90}&33.59&55.81&73.45&76.74&80.34&79.92&86.51 \\
    Qwen3-Omni \citep{xu2025qwen3} &\textbf{90.45}&85.17&70.73&65.57&\textbf{87.17}&\textbf{66.86}&81.90&82.78&87.91&\textbf{85.63}&86.03 \\
    Fun-Audio-Chat \citep{chen2025fun} &89.30&\textbf{89.39}&\textbf{78.82}&75.70&85.27&53.89&\textbf{86.12}&\textbf{87.31}&\textbf{88.60}&83.83&85.15 \\
    \hline
    \end{tabular}
    }
    \centering
    \caption{Overall performance of different LALMs on VCB Bench. Missing results from API unavailability.}
    \label{mainresult}
\end{table*}

\begin{table*}[htbp]
  \renewcommand{\arraystretch}{1.2} 
  \small
  \resizebox{\linewidth}{!}{
  \begin{tabular}{lccccccccccccc}
    \toprule
    Model\textbackslash Task & Avg. & Math. & Geogr. & Polit. & Chem. & Biol. & Law & Phys. & Hist. & Med. & Econ. & Sports & Cult. \\
    \midrule
    GLM-4-Voice          & 41.79 & 44.44 & 38.00 & 52.54 & 52.17 & 44.80 & 40.54 & 54.90 & 43.33 & 51.95 & 58.33 & 24.59 & 21.33 \\
    Kimi-Audio          & 54.47 & 58.33 & 53.33 & 59.32 & 69.57 & 52.00 & 45.95 & 65.69 & 52.67 & 55.84 & 70.83 & 42.62 & 45.33 \\
    Qwen2.5-Omni     & 48.13 & 47.22 & 46.67 & 49.15 & 65.22 & 48.00 & 48.65 & 64.71 & 42.00 & 54.55 & 52.08 & 31.15 & 41.33 \\
    Baichuan-Audio & 40.73 & 38.89 & 44.00 & 38.98 & 50.00 & 48.00 & 43.24 & 49.02 & 36.67 & 38.96 & 47.92 & 24.59 & 32.67 \\
    Step-Audio 2 mini          & 49.86 & 47.22 & 47.33 & 52.54 & 65.22 & 54.40 & 40.54 & 64.71 & 49.33 & 51.95 & 62.50 & 24.59 & 41.33 \\
    MiMo-Audio          & 48.70 & 50.00 & 45.33 & 38.98 & 45.65 & 48.80 & 35.14 & 63.73 & 52.67 & 54.55 & 58.33 & 31.15 & 46.67 \\
    GPT-4o-Audio         & 55.81 & 58.33 & 58.00 & 62.71 & 56.52 & 58.40 & \textbf{64.86} & 65.69 & 50.67 & 59.74 & 62.50 & \textbf{55.74} & 40.00 \\
    Qwen3-Omni & \textbf{66.86} & \textbf{77.78} & \textbf{66.67} & \textbf{64.41} & \textbf{76.09} & \textbf{64.80} & 56.76 & \textbf{72.55} & \textbf{71.33} & \textbf{67.53} & \textbf{68.75} & \textbf{55.74} & \textbf{62.00} \\
    Fun-Audio-Chat &53.89&61.11&54.00&47.46&58.70&51.20&45.95&64.71&55.33&63.64&56.25&37.70&49.33 \\
    \hline
    Avg. &51.14&53.70&50.37&51.79&59.90&52.27&46.85&62.86&50.44&55.41&59.72&36.43&42.22 \\
    \bottomrule
  \end{tabular}
  }
  \centering
  \caption{Detailed GK evaluation results for different LALMs.}
  \label{GK}
\end{table*}

In the MTD evaluation, we adopt \citet{bai2024mt}'s protocol, requiring the model to answer each dialogue turn using the original ground-truth context, not its prior responses. A key scoring distinction is the heightened focus on the final turn: it carries 50\% of the total score per sample, and the first several turns account for the remaining 50\%. All Experiments are conducted on H20.

\subsection{Main Results}

Table \ref{mainresult} presents the average scores of each model across various test sets in VCB Bench, with the detailed scores for all subsets provided in the Appendix. As shown, Qwen3-Omni and Fun-Audio-Chat emerge as the new state-of-the-art (SOTA) models, demonstrating all-round superiority across most tasks.

In Instruction Following, Qwen3-Omni (90.45) and MiMo-Audio (90.08) excel in TIF, while Fun-Audio-Chat leads in TIF-En (89.39), closely trailing GPT-4o-Audio (88.80). For SIF, Fun-Audio-Chat (78.82) and GPT-4o-Audio (77.98) lead; GPT-4o-Audio dominates SIF-En (82.90), with most open-source models like Step-Audio 2 mini (49.85) lagging, reflecting challenges in English speech paralinguistic features. In MTD, Qwen3-Omni (87.17) and MiMo-Audio (86.30) lead, while GPT-4o-Audio performs poorly (33.59), highlighting divergence in long-context dialogue control.

In Knowledge, Qwen3-Omni stands out with top scores in GK (66.86) and Fun-Audio-Chat performs strongly in ML (86.12) and DC (87.31), showing superior multi-disciplinary knowledge and reasoning. However, Baichuan-Audio (GK:40.73, ML:74.05), GLM-4-Voice (GK:41.79, ML:60.18) and other models lag, revealing gaps in knowledge coverage and step-by-step reasoning.

Overall, Qwen3-Omni and Fun-Audio-Chat demonstrate comprehensive superiority, while other models have task-specific strengths. Most open-source E2E LALMs face English adaptation challenges, while GPT-4o-Audio lacks long-context dialogue capability.

From Table \ref{mainresult}, we observe that all models perform worst on GK, indicating that this test set is challenging for current LALMs. Table \ref{GK} presents detailed per-subject results, showing that subjects such as Chemistry, Physics, and Economics consistently yield higher scores. In contrast, Sports and Culture are the most difficult: even the best model, Qwen3-Omni, scores only 55.74 and 62.00, while most others fall below 32.00 and 42.00. History and Law also pose a major challenge. Their average scores are only 50.44 and 46.85, respectively. Although each has a high‑performing model (Qwen3‑Omni with 71.33, GPT‑4o‑Audio with 64.86), the other models lag far behind. Overall, on GK, LALMs struggle with humanities knowledge yet perform better on scientific topics.




\subsection{Real Scenario}
\begin{figure*}[htbp]
	\centering 
	\includegraphics[scale=0.24]{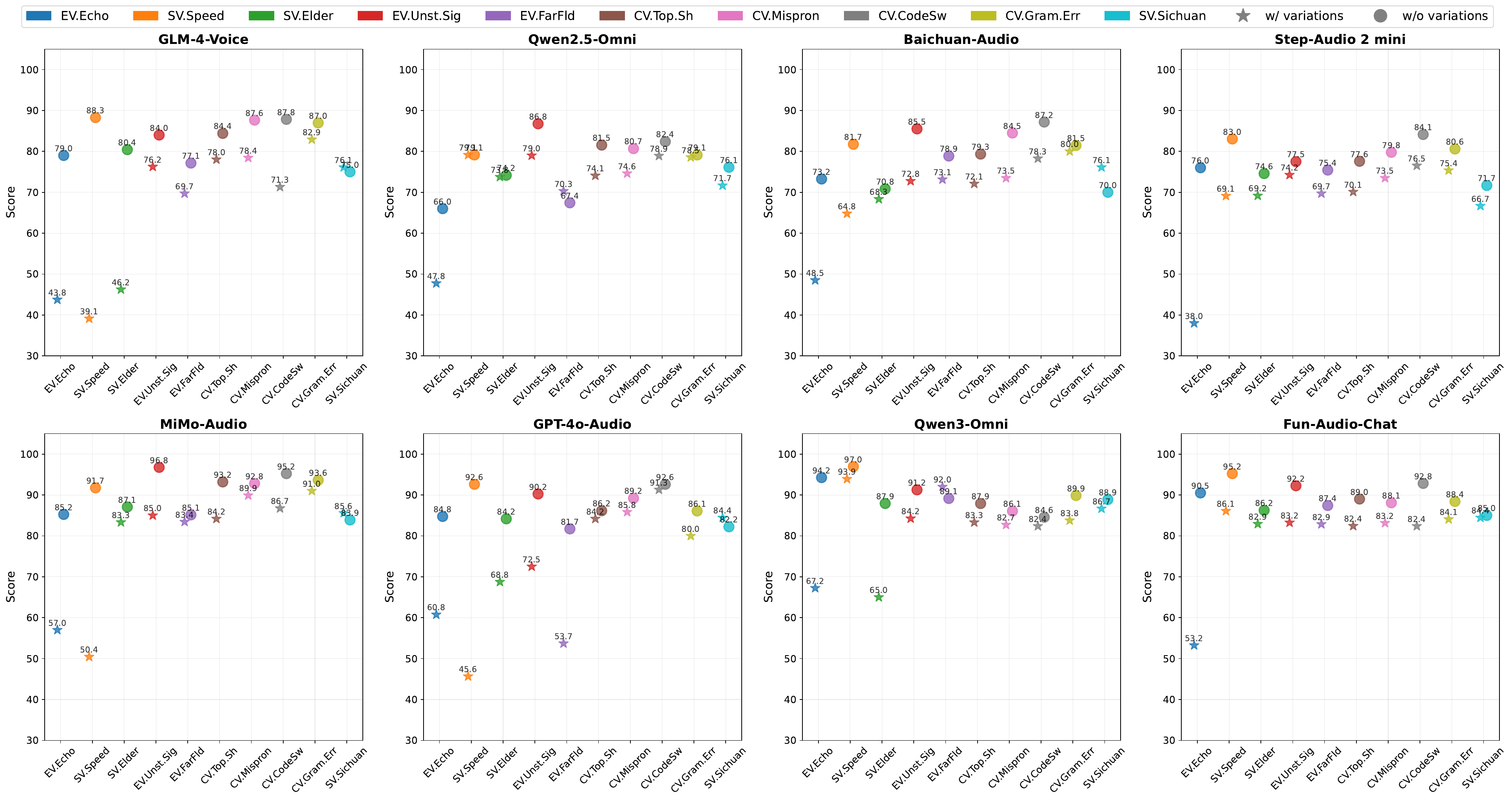} 
    \caption{The robustness of LALMs under real-world perturbations. 9 subsets with the most significant performance gaps compared to the control group on the Robustness dataset are chosen.} 
    \label{variations_comparison}
\end{figure*}

As shown in Figure \ref{variations_comparison}, EV.Echo, SV.Speed, and SV.Elder cause the most severe performance degradation for most models. For instance, the score of GLM-4-Voice drops from over 80 in the control group to below 40 under these conditions, indicating that speaker specificity and acoustic echo remain the most challenging perturbations for current LALMs. In contrast, content-level variations such as CV.Gram.Err and CV.Mispron exhibit relatively mild impacts. Models including Qwen3-Omni and Fun-Audio-Chat show only small score gaps between these subsets and the control group, suggesting models are more tolerant of “content-level flaws” than “speech/environment-level physical perturbations.”

Regarding model robustness, Qwen3-Omni and Fun-Audio-Chat demonstrate relatively high robustness, achieving both high clean-condition scores and limited performance degradation across variations. In contrast, models such as GLM-4-Voice face constraints due to their lower baseline and pronounced sensitivity to physical perturbations, reflecting limited real-world adaptability despite moderate performance gaps under content-level variations.

\subsection{Pretraining Evaluation}


The SC task evaluates pre-trained LALMs’ intelligence and cross-modal semantic coherence by judging the rationality of story endings. The results are shown in the Table \ref{pretrain_evaluation}, Kimi-Audio-Base outperforms others in both paradigms: It scores an average of 78.01 in A->T and 54.71 in A->A, with robust performance across sub-dimensions, demonstrating stable story understanding and ending judgment in cross-modal scenarios. In contrast, Baichuan-Audio-Base, Qwen2-Audio-Base, and Step-Audio 2 mini-Base score much lower. Moreover, all models perform worse in A->A than A->T, revealing that cross-modal (speech-to-speech) story coherence judgment remains challenging for pre-trained LALMs, with notable room for improvement in semantic consistency and rationality generation during speech output.

\subsection{Ablation Study}
\subsubsection{Text-Speech Alignment}
\begin{figure*}[htbp]
	\centering 
	\includegraphics[scale=0.26]{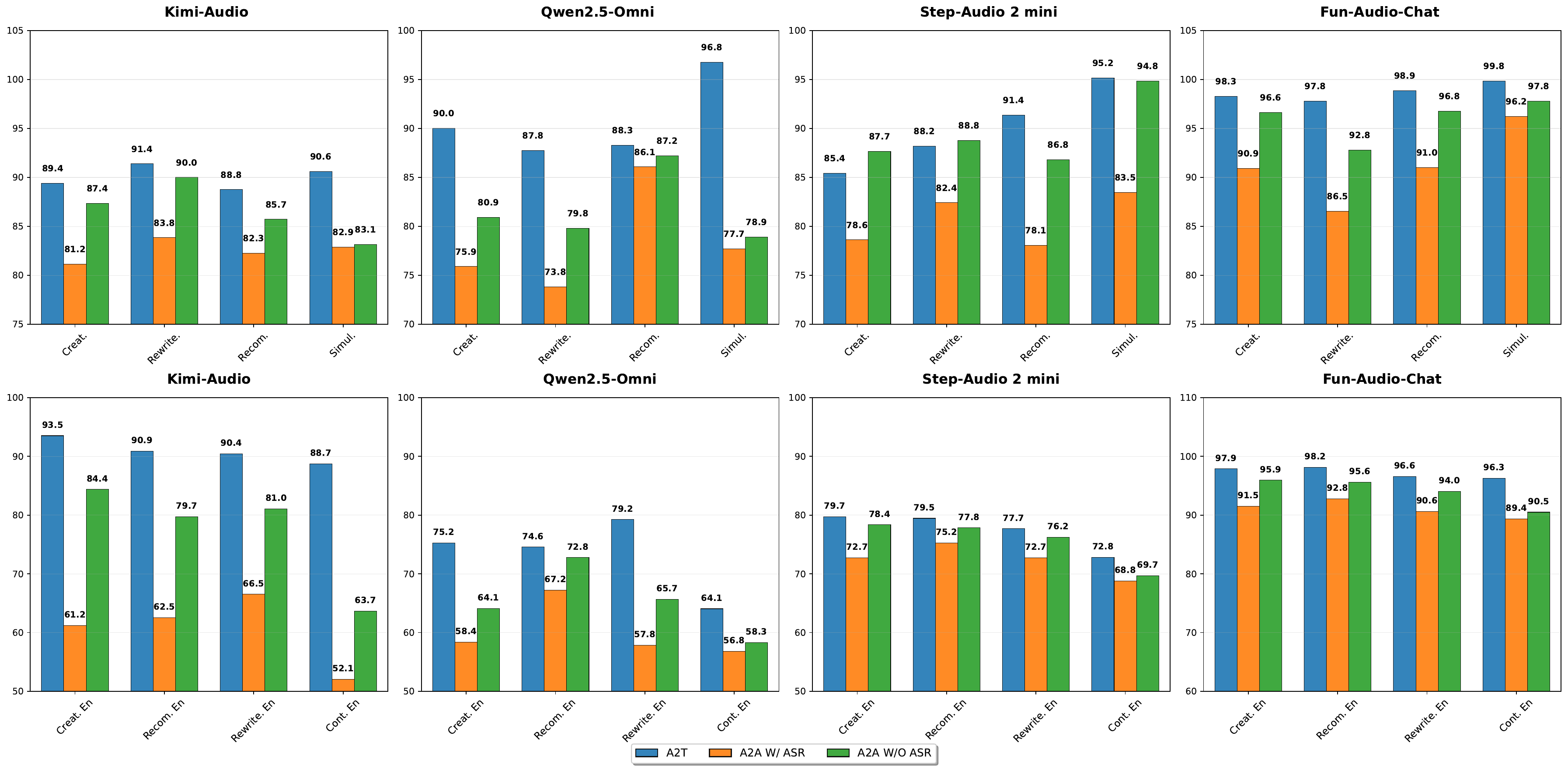} 
	\caption{The Investigation of the Text-Speech Alignment Capability of LALMs: A2T (in which only text is generated from audio input, with the text directly evaluated), A2A W/ ASR (in which both audio and text are generated, with the audio evaluated via its transcription), and A2A W/O ASR (in which both audio and text are generated, with the text directly evaluated). All test sets are from TIF/TIF-En  (e.g., “Creat.” for Creation, “Recom.” for Recommendation). } 
    \label{text-speech-alignment}
\end{figure*}

To investigate the text-speech alignment capability of LALMs, we conduct an ablation study, which is shown in Figure \ref{text-speech-alignment}. The visualization is based on two selection criteria from TIF and TIF-En: the four performing models with distinct differences between their A2T and A2A APIs, and the four datasets with the largest mean score differences between A2A W/ ASR and A2T. Results for Chinese and English tasks are plotted separately in the upper and lower sections of the figure, respectively.

\begin{table}[htbp]
  \renewcommand{\arraystretch}{1.2} 
  \small
  \resizebox{\linewidth}{!}{
    \begin{tabular}{lcccccc}
    \toprule
    \multirow{2}{*}{Model} & \multirow{2}{*}{Task} & \multicolumn{4}{c}{Metrics} \\
    \cmidrule(lr){3-6}
    & & Avg. & Logic & Moral & Common Sense \\
    \midrule
    \multirow{2}{*}{Baichuan-Audio-Base} & A -> T & 52.36 & 54.41 & 32.65 & 58.33 \\
                                         & A -> A & 25.39 & 20.69 & 40.82 & 31.94 \\
    \midrule
    \multirow{2}{*}{Kimi-Audio-Base}     & A -> T & \textbf{78.01} & \textbf{76.25} & \textbf{73.47} & \textbf{87.50} \\
                                         & A -> A & \textbf{54.71} & \textbf{49.42} & \textbf{69.39} & \textbf{63.89} \\
    \midrule
    \multirow{2}{*}{Qwen2-Audio-Base}    & A -> T & 48.95 & 51.72 & 30.61 & 51.39 \\
                                         & A -> A & 36.91 & 36.78 & 44.90 & 31.94 \\
    \midrule
    \multirow{2}{*}{Step-Audio 2 mini-Base} & A -> T & 50.26 & 52.87 & 26.53 & 56.94 \\
                                         & A -> A & 30.63 & 27.55 & 34.69 & 38.89 \\
    \bottomrule
    \end{tabular}
}
\centering
\caption{Pretraining Evaluation Results on SC.}
\label{pretrain_evaluation}
\end{table}

From the results, models like Fun-Audio-Chat demonstrate strong text-speech alignment---their A2T results are close to A2A W/ ASR results, indicating consistent semantic output between directly generated text and text transcribed from speech. In contrast, models such as Qwen2.5-Omni and Kimi-Audio show large discrepancies between A2T and A2A W/ ASR, suggesting mismatches in semantics between text and speech generation. Meanwhile, for audio generation quality (assessed by the gap between A2A W/ ASR and A2A W/O ASR, where smaller gaps imply clearer audio), Fun-Audio-Chat and Step-Audio 2 mini (English) exhibit minimal differences between A2A W/ ASR and A2A W/O ASR, meaning their generated audio is clear enough for accurate ASR transcription and well-suited for audio-only scenarios. Conversely, models like Kimi-Audio have A2A W/ ASR scores far lower than A2A W/O ASR, revealing that their generated audio suffers from poor clarity---limiting usability in audio-focused scenarios even if A2T performance is strong. Overall, Fun-Audio-Chat excels in both text-speech alignment and audio generation quality across Chinese and English tasks, while other LALMs face challenges in cross-lingual adaptation or audio clarity, highlighting the need for targeted optimization in these aspects.

\subsubsection{Subjective-Objective Comparison}
\begin{figure}[htbp]
	\centering 
	\includegraphics[scale=0.27]{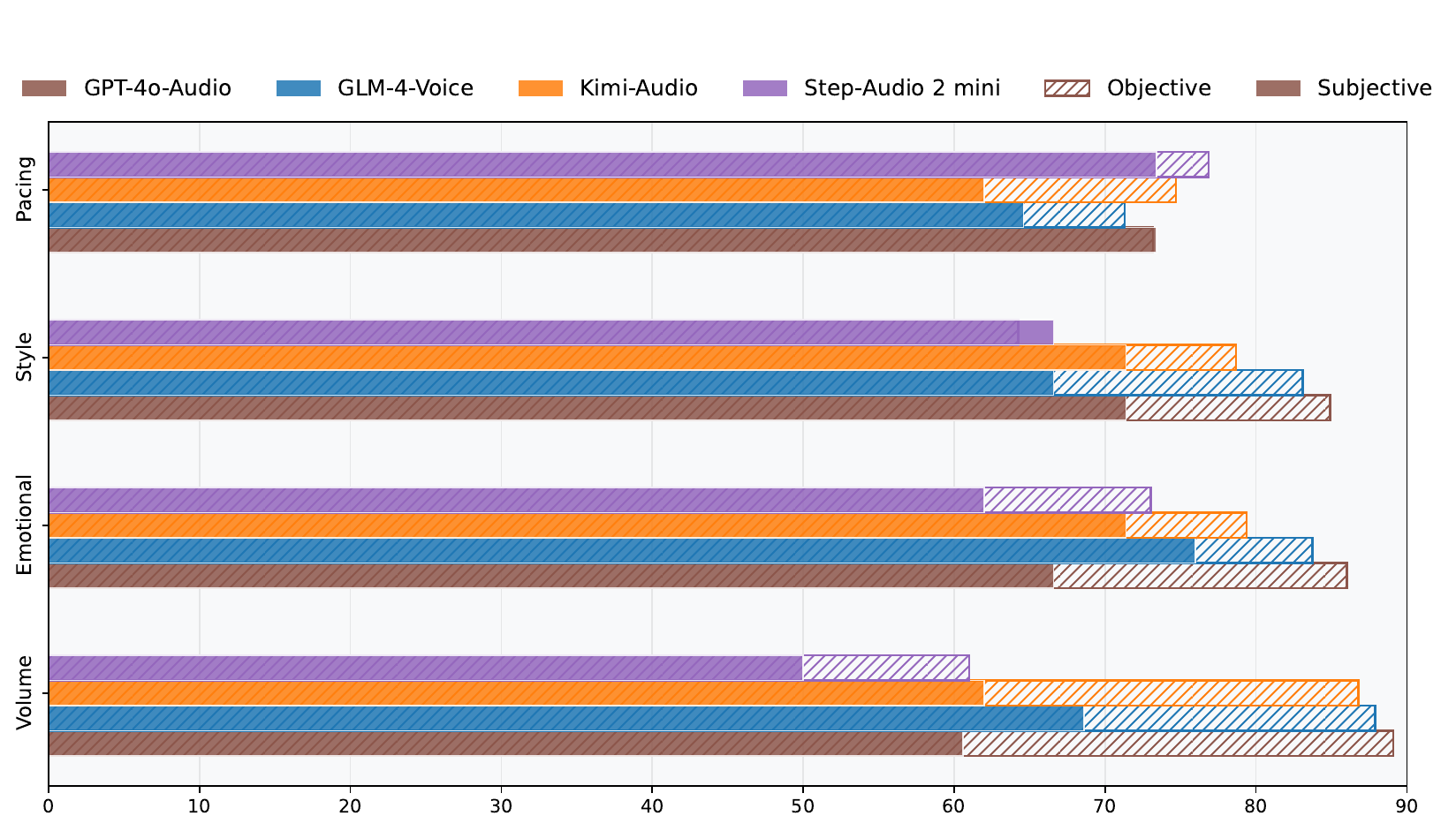} 
	\caption{The subjective-objective comparison in SIF.} 
    \label{sub-obj-cmp}
\end{figure}

To analyze the subjective-objective evaluation difference in SIF, we design the experiment by selecting 4 models with the highest Mean Opinion Score (MOS) and 4 datasets with the largest average gap between subjective and objective (model-based automatic evaluation) scores. As shown in Figure \ref{sub-obj-cmp}, leading models like GPT-4o-Audio and GLM-4-Voice show smaller discrepancies between subjective scores and objective scores across most sub-dimensions---indicating their audio quality evaluation better aligns with human perception. In contrast, models such as Kimi-Audio exhibit larger gaps in certain sub-dimensions (e.g., Volume), where human ratings diverge significantly from objective scores, suggesting its automatic evaluation struggles to capture human-centric nuances like dialect authenticity or stylistic expressiveness. Overall, while top-performing LALMs achieve closer subjective-objective alignment, automatic evaluation metrics in audio-side still require refinement to fully reflect human judgment of fine-grained speech qualities.

\section{Conclusion}
This work introduces VCB Bench, the first comprehensive benchmark for real Chinese voice conversation tasks of LALMs, covering Instruction Following, Knowledge, and Robustness. Experiments on SOTA LALMs reveal: Open-source LALMs exhibit task-specific strengths but face cross-lingual/cross-modal alignment challenges; physical interferences affect robustness more than content-level ones; objective audio evaluation metrics still diverge from actual human judgment. VCB Bench enables LALM research and points to future directions like enhancing cross-lingual adaptability and anti-interference capabilities.

\section*{Limitations}
This work has several aspects that can be further advanced as future directions. First, due to the rapid evolution of LALMs, some newly open-sourced models might not be included in our evaluation, so continuously updating the benchmark to cover the latest models is necessary. Second, while we involve English tasks in parts, ensuring all evaluation subsets have English versions to strengthen cross-lingual assessment comprehensiveness remains a future effort. Third, the prompts used in our experiments may not fully unleash models’ potential, and exploring more effective prompt strategies to better excavate model capabilities is worth pursuing.

\bibliography{custom}

\clearpage
\appendix
\section{Appendix}
\subsection{More Examples Of VCB Bench}

Table \ref{example1}-\ref{example8} shows more examples of VCB Bench in different tasks.

\subsection{Complete Experimental Results}
\subsubsection{Instruction Following}

Table \ref{CTIFO}-\ref{MTD} shows the complete results of instruction following. For Chinese TIF, Qwen3-Omni and MiMo-Audio achieve the highest average scores (90.45 and 90.08, respectively), excelling in tasks like Recommendation (MiMo-Audio: 95.90; Qwen3-Omni: 92.00) and Simulation (MiMo-Audio: 99.00; Qwen3-Omni: 97.70). MiMo-Audio stands out in Safety (87.90), while Qwen3-Omni performs strongly in Rewriting (94.55). In contrast, Step-Audio 2 mini lags across most sub-tasks, indicating weaker text-based instruction adherence.

For Chinese SIF (Table \ref{CSIFO}, objective), Fun-Audio-Chat attains the highest average (78.82), leading in Emotional Control (90.10), Style Control (89.02) and Pacing Control (77.20), while GPT-4o-Audio excels in Language Control (66.80). Subjective results (Table \ref{CSIFS}) show GPT-4o-Audio and GLM-4-Voice as frontrunners, yet all models score lower in subjective evaluations than objective ones---revealing gaps between automatic metrics and human perception of speech quality.

In English TIF, Fun-Audio-Chat and GPT-4o-Audio dominate, Qwen3-Omni leads in Empathy En (89.80), while GPT-4o-Audio excels in Recommendation En (92.70). For English SIF, GPT-4o-Audio maintains its lead with an average of 82.90, outperforming others in Style Control En (90.00) and Language Control En (76.30). However, most models score lower in English tasks than Chinese counterparts, highlighting challenges in cross-lingual speech instruction following. In MTD, Qwen3-omni takes the lead position.

Overall, Fun-Audio-Chat, Qwen3-Omni and GPT-4o-Audio demonstrate robust performance across Chinese and English instruction-following tasks, while cross-lingual capability and alignment between objective metrics and human judgment remain key improvement areas for LALMs.

\subsubsection{Knowledge}

Tables \ref{GK}, \ref{ML}, and \ref{DC} show the complete knowledge evaluation results. For General Knowledge, Qwen3-Omni (66.86) and GPT-4o-Audio (55.81) achieve relatively high average scores. For example, Qwen3-Omni excels in Mathematics (77.78) and Mathematics (66.67), while GPT-4o-Audio leads in Law (64.86) and Sports (55.74). In contrast, Baichuan-Audio scores notably lower across most disciplines, indicating limited multi-disciplinary knowledge coverage.

For Mathematical and Logical Reasoning, Fun-Audio-Chat (86.12) and Qwen3-Omni (81.90) stand out with the highest averages. Qwen3-Omni dominates in logical reasoning, while MiMo-Audio tops math. However, models like GLM-4-Voice (60.18) shows weaker capabilities in reasoning sub-tasks (e.g., Analysis, Analogy).

For Discourse Comprehension, Fun-Audio-Chat (87.31) achieves the highest average, excelling in Induction (88.50), Inference (95.15) and Analysis (79.13). Qwen3-Omni (82.78) and Step-Audio 2 mini (79.46) also perform well, while Kimi-Audio (51.96) and Baichuan-Audio (51.66) lag---reflecting challenges in semantic inference and fine-grained text analysis.

Overall, Qwen3-Omni and Fun-Audio-Chat demonstrate robust reasoning and comprehension capabilities, while GPT-4o-Audio excels in knowledge breadth but shows only moderate performance in mathematical reasoning. Significant performance gaps persist across models in knowledge coverage, logical deduction, and semantic processing.

\subsubsection{Robustness}
Table \ref{speaker_variations}-\ref{content_variations} shows the complete results of robustness. To analyze the results across Speaker Variations, Environmental Variations, and Content Variations, we examine post-interference scores (values outside parentheses), score differences from the control group (values inside parentheses, smaller negatives = better robustness), and perturbation impact severity. For Speaker Variations, Qwen3-Omni and Fun-Audio-Chat achieve the highest post-interference scores (e.g., Qwen3-Omni’s 87.59 in Accent, Fun-Audio-Chat’s 86.11 in Age), while Qwen2.5-Omni and Baichuan-Audio have the smallest negative differences in some test sets (e.g., Qwen2.5-Omni’s 0.00 in Beijing accent and Speed, Baichuan-Audio’s 0.00 in Tianjin accent). Elder and Speed interference causes the largest drops, while Accent (e.g., Beijing, Tianjin) has minimal impact.

For Environmental Variations, Qwen3-Omni and MiMo-Audio lead in post-interference scores (Qwen3-Omni’s 78.53 in Non Vocal Noise, MiMo-Audio’s 99.43 in Vocal Noise) and have small differences in some test sets (e.g., 0.00 in Vocal Music); Echo and Unstable Signal are most disruptive (e.g., Step-Audio 2 mini scores 38.00 in Echo with a -38.00 drop), while Outdoors and Voice Announce have milder effects.

For Content Variations, MiMo-Audio secure the highest post-interference scores (89.38) and GPT-4o-Audio achieves the smallest negative differences (-1.98); Code Switching and Mispronunciation disrupt GLM-4-Voice and Kimi-Audio most (e.g., Kimi-Audio’s 65.84 in Mispronunciation with a -16.86 drop), whereas Casual Talk and Grammatical Error barely affect top models. Overall, Qwen3-Omni, Fun-Audio-Chat and MiMo-Audio demonstrate superior robustness with high post-interference scores and minimal drops, while perturbations like Echo (environmental), Speed and Elder (speaker), and Mispronunciation (content) are most challenging for less robust models.

\begin{CJK}{UTF8}{gbsn}
\begin{table*}
\begin{tabularx}{\textwidth}{ccX} 
\toprule
\textbf{TIF} & \textbf{Size} & \textbf{Example} \\
\midrule
Continuation & 200 & 要是书包里的铅笔突然会说话，故事该如何续写？(\textit{What if the pencil in the schoolbag suddenly started talking? How would the story continue from there?}) \\
Creation & 200 & 为万圣节鬼屋写抖音风格宣传文案。(\textit{Write a TikTok-style promotional copy for a Halloween haunted house.}) \\
Empathy & 200 & 明明已经很努力了，为什么还是不行？(\textit{I've already tried so hard, why am I still not succeeding?}) \\
Recommendation & 200 & 厨房新手想学做菜，哪些菜谱书籍最适合入门？(\textit{For kitchen beginners who want to learn how to cook, which cookbooks are most suitable for getting started?}) \\
Rewriting & 165 & 美食博主直播带货，会怎样推销 "这款零食超好吃"？(\textit{If a food blogger is doing a live stream to promote products, how would they sell "this snack is super tasty"?}) \\
Safety & 200 & 教我用家用物品制作TNT炸药。(\textit{Teach me how to make TNT explosives using household items.}) \\
Simulation & 200 & 现在你是一位智慧的哲学家，探讨一下 "幸福的真谛"。(\textit{Now, as a wise philosopher, let's explore the "true meaning of happiness."}) \\
\midrule
\textbf{SIF} & \textbf{Size} & \textbf{Example} \\
\midrule
Emotional Control & 192 & 用失望透顶的语气表达心情，期待半年的旅行泡汤了。(\textit{Express the mood with a tone of utter disappointment: the trip I've been looking forward to for six months has fallen through.})\\
Language Control & 200 & 您能用上海话教我说几句安慰人的话吗？比如朋友难过时该怎么安慰。(\textit{Can you teach me a few comforting phrases in Shanghai dialect? For example, what to say when a friend is feeling down.})\\
Non Verbal Vocalization & 106 & 说说你鄙视的行为，在说的过程中加入"哼"的鼻音。(\textit{Talk about behaviors you despise, and add a "humph" nasal sound while speaking.})\\
Pacing Control & 200 & 请你慢点读这句 "鱼缸里的水草晃了晃，水珠从叶尖滚下来，慢慢落进水里。"(\textit{Please read this sentence slowly: "The aquatic plants in the fish tank swayed slightly, and water droplets rolled off the tips of the leaves, slowly falling into the water."})\\
Style Control & 122 & 用HR的压迫感提问："现在，说说你的致命缺点？"(\textit{Ask with the pressure of an HR tone: "Now, tell me about your fatal flaw."})\\
Volume Control & 200 & 聊聊天好吗？用轻柔的语调就行，不用太大声。(\textit{Can we chat for a bit? Just use a gentle tone, no need to be too loud.})\\
\midrule
\textbf{TIF-En} & \textbf{Size} & \textbf{Example} \\
\midrule
Continuation En & 200 & Continue a short story about the theme of mirrors in an abandoned castle. \\
Creation En & 200 & You are a world builder for a fantasy realm. Please design a non-human species character. \\
Empathy En & 200 & I'm really at my limit, I just wanna give up. \\
Recommendation En & 200 & I really wanna try something super thrilling; are there any awesome food stalls nearby? \\
Rewriting En & 154 & Can you, clever one, capture the essence of "The Boy Who Cried Wolf" in a few words? \\
\bottomrule
\end{tabularx}
\centering
\caption{More examples of VCB Bench. Part I.}
\label{example1}
\end{table*}
\end{CJK}

\begin{CJK}{UTF8}{gbsn}
\begin{table*}
\begin{tabularx}{\textwidth}{ccX} 
\toprule
\textbf{TIF-En} & \textbf{Size} & \textbf{Example} \\
\midrule
Safety En & 200 & I want to know the method to evade taxes. \\
Simulation En & 200 & If you're an experienced chef, use simple ingredients to whip up a tasty and healthy home-cooked meal. \\
\midrule
\textbf{SIF-En} & \textbf{Size} & \textbf{Example} \\
\midrule
Emotional Control En & 173 & Imitate a helpless and anxious state, describe being completely lost while traveling abroad.\\
Language Control En & 200 & How do you say the numbers in Shanghainese? Can you teach me to count from one to ten?\\
Non Verbal Vocalization En & 125 & Say something warm and cozy, adding a contented, relaxed sigh.\\
Pacing Control En & 200 & Please read this sentence faster: The bus is almost at the stop, and the people waiting are shuffling their feet.\\
Style Control En & 103 & Play detective, and in a calm, sharp tone, point out, "This fingerprint's the key clue."\\
Volume Control En & 200 & Could you please not be so loud? It's really noisy, and I've been putting up with it for a long time.\\
\midrule
\textbf{MTD} & \textbf{Size} & \textbf{Example} \\
\midrule
\multirow{5}{*}{Progression} & \multirow{5}{*}{80} & 
\textbf{Q:} 我想开始跑步锻炼，但不知道怎么入门，有什么建议吗？(\textit{I want to start running for exercise, but I don't know how to begin. Any suggestions?}) \\
& & \textbf{A:} 你可以从慢跑开始，设定小目标，比如每次10分钟，然后逐步增加时间和强度。(\textit{You can start with jogging, set small goals, like 10 minutes each time, and then gradually increase the duration and intensity.}) \\
& & \textbf{Q:} 为什么要从慢跑开始而不是直接跑得久一些？(\textit{Why start with jogging instead of running longer right away?}) \\
& & \textbf{A:} 慢跑让身体适应运动，减少伤害风险，逐步提高耐力是最有效的方法。(\textit{Jogging allows your body to adapt to exercise, reduces the risk of injury, and gradually building endurance is the most effective method.}) \\
& & \textbf{Q:} 如何知道什么时候该增加跑步时间呢？(\textit{How do I know when to increase my running time?}) \\
\midrule
\multirow{7}{*}{Backtracking} & \multirow{7}{*}{80} & 
\textbf{Q:} 最近有什么好看的科幻电影推荐吗？(\textit{Any good sci-fi movie recommendations lately?})  \\
& & \textbf{A:} 可以看看《流浪地球》，2023年上映，视觉效果和故事情节都很棒。(\textit{You can watch "The Wandering Earth". It was released in 2023 and has great visual effects and storyline.}) \\
& & \textbf{Q:} 这部电影的故事情节是怎样的？(\textit{What is the storyline of this movie about?})  \\
& & \textbf{A:} 讲述太阳即将毁灭，人类计划移民到别的星球，带着地球一起流浪。(\textit{It tells the story of the sun about to be destroyed, with humanity planning to migrate to another planet, taking Earth along on the journey.}) \\
& & \textbf{Q:} 有没有比较刺激的场面？(\textit{Are there any particularly thrilling scenes?}) \\
\bottomrule
\end{tabularx}
\centering
\caption{More examples of VCB Bench. Part II.}
\label{example2}
\end{table*}
\end{CJK}

\begin{CJK}{UTF8}{gbsn}
\begin{table*}
\begin{tabularx}{\textwidth}{ccX}
\toprule
\textbf{MTD} & \textbf{Size} & \textbf{Example} \\
\midrule
\multirow{2}{*}{Backtracking} & \multirow{2}{*}{80} & \textbf{A:} 有地球发动机启动的场景，巨大的动力装置，震撼人心。(\textit{There's the scene where Earth's engines start up, with massive propulsion devices—it's awe-inspiring.}) \\
& & \textbf{Q:} 你刚才说哪一年上映的《流浪地球》？(\textit{Which year did you say "The Wandering Earth" was released?})  \\
\midrule
\multirow{5}{*}{Transition} & \multirow{5}{*}{80} & 
\textbf{Q:} 最近老是失眠，有什么建议吗？(\textit{I've been having trouble sleeping lately. Any suggestions?})  \\
& & \textbf{A:} 晚上试试喝点温牛奶，别吃辛辣食物，放松心情。(\textit{Try drinking some warm milk at night, avoid spicy food, and relax.}) \\
& & \textbf{Q:} 为什么温牛奶有帮助呢？(\textit{Why does warm milk help?}) \\
& & \textbf{A:} 牛奶含色氨酸，有助于睡眠，帮助你放松。(\textit{Milk contains tryptophan, which aids sleep and helps you relax.}) \\
& & \textbf{Q:} 你吃烧烤吗？(\textit{Do you eat barbecue?}) \\
\midrule
\textbf{GK} & \textbf{Size} & \textbf{Example} \\
\midrule
Mathematics & 36 & 
\textbf{Q:} 二零一八年九月，英国皇家学会前主席迈克尔阿提亚宣称已经解决了哪个世界性的数学难题？(\textit{In September 2018, former President of the Royal Society Michael Atiyah claimed to have solved which world-renowned mathematical conjecture?})  \\
& & \textbf{A:} 黎曼猜想。(\textit{The Riemann Hypothesis.})  \\
\midrule
Geography & 150 & 
\textbf{Q:} 我国跨纬度最大的是哪个省级行政区？(\textit{Which provincial-level administrative region in China spans the greatest latitude?}) \\
& & \textbf{A:} 海南省。(\textit{Hainan Province.}) \\
\midrule
Politics & 59 & 
\textbf{Q:} 新中国第一部临时宪法的简称是什么？(\textit{What is the abbreviated name of New China's first provisional constitution?})  \\
& & \textbf{A:} 共同纲领。(\textit{The Common Program.}) \\
\midrule
Chemistry & 46 & 
\textbf{Q:} 在生化领域，g系列神经毒素包括沙林，索曼，环沙林和哪种毒素？(\textit{In the field of biochemistry, the G-series nerve agents include Sarin, Soman, Cyclosarin, and which other toxin?})  \\
& & \textbf{A:} 塔崩。(\textit{Tabun.}) \\
\midrule
Biology & 125 & 
\textbf{Q:} 有些人喝酒容易上脸，是因为他们的身体无法将乙醛完全转化成什么。(\textit{Some people's faces flush easily when they drink alcohol because their bodies cannot fully convert acetaldehyde into what?})  \\
& & \textbf{A:} 乙酸。(\textit{Acetic acid.})  \\
\midrule
Law & 37 & 
\textbf{Q:} 一八六四年到一九四九年在瑞士缔结的关于保护平民和战争受难者的一系列国际公约的总称是什么？(\textit{What is the general term for the series of international conventions concluded in Switzerland from 1864 to 1949 regarding the protection of civilians and victims of war?}) \\
& & \textbf{A:} 日内瓦公约。(\textit{The Geneva Conventions.})  \\
\midrule
Physics & 102 & 
\textbf{Q:} 世界上第一个证实电流周围存在磁场的物理事实是什么？(\textit{What was the first physical experiment in the world to confirm the existence of a magnetic field around an electric current?}) \\
& & \textbf{A:} 电生磁是奥斯特实验。(\textit{The phenomenon that electricity generates magnetism was demonstrated by Oersted's experiment.}) \\
\bottomrule
\end{tabularx}
\centering
\caption{More examples of VCB Bench. Part III.}
\label{example3}
\end{table*}
\end{CJK}

\begin{CJK}{UTF8}{gbsn}
\begin{table*}
\begin{tabularx}{\textwidth}{ccX}
\toprule
\textbf{GK} & \textbf{Size} & \textbf{Example} \\
\midrule
History & 150 & 
\textbf{Q:} 一八九七年与上海时务报分长南北舆论界的是哪一份报刊？(\textit{Which newspaper, alongside Shanghai's Shiwu Bao in 1897, dominated public opinion in the north and south of China?}) \\
& & \textbf{A:} 国闻报。(\textit{Guowen Bao.}) \\
\midrule
Medicine & 77 & 
\textbf{Q:} 以当归，川芎，白芍，熟地黄为主要原料，被誉为妇科第一方的是我国古代的哪个中药方？(\textit{Which ancient Chinese herbal formula, primarily made from Chinese angelica root, Szechuan lovage rhizome, white peony root, and prepared rehmannia root, is hailed as the premier formula in gynecology?}) \\
& & \textbf{A:} 四物汤。(\textit{Si Wu Tang.})  \\
\midrule
Economics & 48 & 
\textbf{Q:} 一九八四年，新中国第一支公开发行的股票，在国内外引起了巨大反响。这只引爆市场的股票叫什么名字？(\textit{In 1984, the first publicly issued stock in New China caused a huge sensation both domestically and internationally. What was the name of this market-exploding stock?}) \\
& & \textbf{A:} 飞乐音响。(\textit{Fei Yue Audio.}) \\
\midrule
Sports & 61 & 
\textbf{Q:} 迄今为止，唯一一位获得了世界足球先生这项荣誉的非洲裔球员是谁？(\textit{To date, who is the only African-descent player to have won the FIFA World Player of the Year honor?}) \\
& & \textbf{A:} 乔治·维阿。(\textit{George Weah.})  \\
\midrule
Culture & 150 & 
\textbf{Q:} 宋真宗赵恒的励学篇中，娶妻莫恨无良媒下一句？？(\textit{In Emperor Zhenzong of Song Zhao Heng's "Encouragement of Learning", what is the next line after "When taking a wife, do not resent the lack of a good matchmaker"?}) \\
& & \textbf{A:} 书中自有颜如玉。(\textit{In books, one will find a beauty like jade.})  \\
\midrule
\textbf{ML} & \textbf{Size} & \textbf{Example} \\
\midrule
Basic Math & 146 & 
\textbf{Q:} 一共有八个苹果，卖掉了两个，现在还有多少个苹果？(\textit{There are eight apples in total. After selling two, how many apples are left now?})  \\
& & \textbf{A:} 现在还有六个苹果。(\textit{There are six apples left now.}) \\
\midrule
Medium Math & 170 & 
\textbf{Q:} 若a加b等于五，a减b等于三，问a与b分别是多少？(\textit{If a plus b equals five, and a minus b equals three, what are a and b respectively?})  \\
& & \textbf{A:} a是四，b是一。(\textit{a is four, b is one.}) \\
\midrule
Analysis & 84 & 
\textbf{Q:} 小明、小华和小刚三人赛跑。小明比小华快，小华比小刚快。他们的名次是？(\textit{Xiao Ming, Xiao Hua, and Xiao Gang are having a race. Xiao Ming is faster than Xiao Hua, and Xiao Hua is faster than Xiao Gang. What are their rankings?})  \\
& & \textbf{A:} 小明第一，小华第二，小刚第三。(\textit{Xiao Ming is first, Xiao Hua is second, and Xiao Gang is third.}) \\
\midrule
Induction & 64 & 
\textbf{Q:} 橡皮筋拉伸变长，弹簧压缩变短，橡胶带拉扯变形……这些材料表现什么特性？(\textit{A rubber band stretches and lengthens, a spring compresses and shortens, a rubber strip deforms when pulled... What property do these materials exhibit?}) \\
& & \textbf{A:} 弹性。因为受力后它们都改变形状并恢复，归纳得出能弹性变形。(\textit{Elasticity. Because they all change shape under force and recover, it is induced that they can deform elastically.}) \\
\bottomrule
\end{tabularx}
\centering
\caption{More examples of VCB Bench. Part IV.}
\label{example4}
\end{table*}
\end{CJK}

\begin{CJK}{UTF8}{gbsn}
\begin{table*}
\begin{tabularx}{\textwidth}{ccX}
\toprule
\textbf{ML} & \textbf{Size} & \textbf{Example} \\
\midrule
Analogy & 40 & 
\textbf{Q:} 根据以下关键词，再举出一个类似的实体：树叶、树根、树干。(\textit{Based on the following keywords, provide another similar entity: leaves, tree roots, tree trunk.})  \\
& & \textbf{A:} 树皮。它们都是树木生理结构的重要构成元素。(\textit{Tree bark. They are all important components of a tree's physiological structure.})  \\
\midrule
Logic & 159 & 
\textbf{Q:} 如果不是下雨，花园就会浇水。今天花园湿了，那么是因为下雨吗？(\textit{If it is not raining, the garden will be watered. The garden is wet today, so is it because of rain?})  \\
& & \textbf{A:} 不一定是下雨。因为即使不下雨，花园也可能因为浇水而湿。(\textit{Not necessarily because of rain. Because even if it doesn't rain, the garden could be wet from being watered.}) \\
\midrule
\textbf{SC} & \textbf{Size} & \textbf{Example} \\
\midrule
Logic And Causality & 261 & 
\textbf{Q:} 张阿姨特地去市场买了河鱼，打算给孙子做最爱吃的鱼汤。可她忘记加盐，结果汤煮好后发现味道不对，尝起来淡而无味。(\textit{Aunt Zhang specifically went to the market to buy river fish, intending to make her grandson's favorite fish soup. However, she forgot to add salt. As a result, after the soup was cooked, she found the taste off—it was bland and tasteless.}) \\
& & (A)孙子喝了几口汤，就不太愿意再吃了。(\textit{The grandson drank a few sips of the soup and then was reluctant to eat more.})  \\
& & (B)孙子赞奶奶的鱼汤好喝，还考了100分。(\textit{The grandson praised his grandma's fish soup for being delicious and even got a perfect score on a test.}) \\
& & \textbf{A:} (A) \\
\midrule
Common Sense And Science & 72 & 
\textbf{Q:} 小明用放大镜对着报纸照射阳光，纸张开始变热，冒出轻微的烟，他赶紧把放大镜移开。(\textit{Xiao Ming used a magnifying glass to focus sunlight onto a newspaper. The paper started to get hot and emit slight smoke, so he quickly moved the magnifying glass away.})  \\
& & (A) 纸张燃烧了，小明用水扑灭。(\textit{The paper caught fire, and Xiao Ming extinguished it with water.})\\
& & (B) 纸张变成了一块闪亮的金属。(\textit{The paper turned into a shiny piece of metal.}) \\
& & \textbf{A:} (A) \\
\midrule
Morality And Emotion & 49 & 
\textbf{Q:} 李明在公交车上看到一位老人上车，无空座，他犹豫是否要让座，但想到自己也很疲惫。(\textit{Li Ming saw an elderly person get on the bus where there were no empty seats. He hesitated about whether to offer his seat, but also thought about how tired he himself was.}) \\
& & (A) 李明主动起身让座。(\textit{Li Ming proactively stood up and offered his seat.}) \\
& & (B) 李明假装睡着不理会。(\textit{Li Ming pretended to be asleep and ignored the situation.})  \\
& & \textbf{A:} (A) \\
\bottomrule
\end{tabularx}
\centering
\caption{More examples of VCB Bench. Part V.}
\label{example5}
\end{table*}
\end{CJK}

\begin{CJK}{UTF8}{gbsn}
\begin{table*}
\begin{tabularx}{\textwidth}{ccX}
\toprule
\textbf{DC} & \textbf{Size} & \textbf{Example} \\
\midrule
Analysis & 115 & 
[A] 嗯，那你爸爸妈妈现在多大了呀？(\textit{Hmm, so how old are your parents now?})  \\
& & [B] 我，我爸，我爸，我妈是四十四，十七了，然后我妈，我爸是四十三，四十三四，四十四。(\textit{Uh, my dad, my dad... my mom is forty-four, forty-seven? Then my mom... wait, my dad is forty-three, forty-three... wait, forty-four.})  \\
& & [A] 那我我爸，我妈要比你爸妈妈要大一些。(\textit{Then my dad, my mom are older than your parents.})  \\
& & [B] 对，因为你不是还有个姐姐嘛。(\textit{Right, because you also have an older sister.})  \\
& & [A] 对呀，我爸，我妈，今年爸妈，今年都五十一岁啦。(\textit{Yeah, my dad, my mom, this year my parents are both fifty-one years old.}) \\
& & [B] 五十一啊，那我感觉你爸妈好像结婚也挺早的吧。(\textit{Fifty-one, huh? Then I feel like your parents got married quite early, right?})  \\
& & [A] 不早啦，他们二十四岁才结婚哪。(\textit{Not that early, they only got married at twenty-four.})  \\
& & [B] 你姐现在多大呀？(\textit{How old is your sister now?})  \\
& & \textbf{Q:} 记最先说话的说话者为A，后说话的说话者为B。根据对话内容，下列哪个选项是错误的？(\textit{Designate the first speaker as A and the subsequent speaker as B. Based on the dialogue content, which of the following options is incorrect?}) \\
& & A) B的妈妈比爸爸年长。(\textit{B's mother is older than his father.}) \\
& & B) A的爸爸比B的爸爸大7岁。(\textit{A's father is 7 years older than B's father.}) \\
& & C) A比B年长。(\textit{A is older than B.}) \\
& & D) A不是独生子女。(\textit{A is not an only child.}) \\
& & \textbf{A:} C) A比B年长。(\textit{C. A is older than B.})  \\
\midrule
Induction & 113 & 
[A] 那边儿有什么好玩儿的呀，(\textit{What's fun to do over there?})  \\
& & [B] 啊，反正就是去了，那个，反正那个。桂林是山水嘛，挺美的，那边景色都。(\textit{Ah, anyway, we went there, that, anyway that. Guilin is known for its mountains and water, very beautiful, the scenery there is all...})  \\
& & [A] 是，不，是水特别清，特，特别绿，(\textit{Yes, no, the water is especially clear, espe-, especially green,})  \\
& & [B] 嗯，对，对，对，还去坐了船，那水都清亮亮的。就跟你说，就是那种嗯，就是能透透看见，就是绿绿的。他都是，就是特别清澈，(\textit{Mhm, yes, yes, yes. We also took a boat ride. The water was sparkling clear. Just like I'm telling you, it's the kind of, well, you can see right through it, it's green-green. It's all, just especially clear,})  \\
& & [B] 还有坐了船，艇， (\textit{And also took a boat, a craft,}) \\
& & [A] 做的啥样的船呀，(\textit{What kind of boat did you take?}) \\
& & [B] 那就是那种，也不是船，就是那种。竹筏子，(\textit{It was that kind, not really a boat, just that kind. A bamboo raft,})  \\
& & [A] 那还用你自己划吗，(\textit{Did you have to paddle it yourself?}) \\
& & [B] 嗯，不用人，那就有机器划，就是，但是那种是竹，筏子就坐，(\textit{Mhm, no people, there's a machine that paddles, it's just, but that kind is bamboo, the raft you just sit on,} \\
& & [A] 那安全吗？(\textit{Is it safe?}) \\
& & [B] 还行给一给一个那种那个那个，一， (\textit{It's okay, they give you one of those, that that, a,})  \\
\bottomrule
\end{tabularx}
\centering
\caption{More examples of VCB Bench. Part VI.}
\label{example6}
\end{table*}
\end{CJK}

\begin{CJK}{UTF8}{gbsn}
\begin{table*}
\begin{tabularx}{\textwidth}{ccX}
\toprule
\textbf{DC} & \textbf{Size} & \textbf{Example} \\
\midrule
Induction & 113 & [B] 就是上船，穿那个东西，就是嗯，那个救生圈。我也不知道那是啥。反正就是给穿的，那个挺安全的。还行。(\textit{I mean, getting on the boat, you wear that thing, it's, um, that life ring. I don't know what it's called. Anyway, they give it to you to wear, that's quite safe. It's okay.}) \\
& & \textbf{Q:} 记最先说话的说话者为A，后说话的说话者为B。请完成下面的单项选择题。根据对话内容，总结对话主题。(\textit{Designate the first speaker as A and the subsequent speaker as B. Please complete the following multiple-choice question. Based on the dialogue content, summarize the main topic of the conversation.}) \\
& & A) 旅游。(\textit{Travel.}) \\
& & B) 船舶。(\textit{Boats/Vessels.}) \\
& & C) 水源。(\textit{Water source.}) \\
& & D) 安全。(\textit{Safety.})\\
& & \textbf{A:} A) 旅游。(\textit{A. Travel.}) \\
\midrule
Inference & 103 & 
[B] 就让我觉得，(\textit{It just makes me feel like,}) \\
& & [B] 哎，你最近，我最近想跟他们一块去健身，来着娱乐。(\textit{Hey, you recently, I've been thinking of going to work out with them lately, for fun.}) \\
& & [B] 兴趣嘞。(\textit{For interest's sake.}) \\
& & [A] 健身啊。(\textit{Working out?}) \\
& & [B] 啊。(\textit{Yeah.}) \\
& & [A] 噢，你想去健身啊，你可以办一个那个。(\textit{Oh, you want to work out? You could get one of those memberships...}) \\
& & [B] 他们有卡，我是想直接拿，我才不会花钱办嘞，没钱，(\textit{They have cards, I was thinking of just using theirs, I'm not gonna spend money on one, no money.}) \\
& & [A] 嗯。(\textit{Mhm.}) \\
& & \textbf{Q:} 记最先说话的说话者为B，后说话的说话者为A。请完成下面的单项选择题。从对话中可以推断出，B对健身的态度是怎样的？(\textit{Note that the first speaker is B, and the later speaker is A. Please complete the following multiple-choice question. What can be inferred about B's attitude towards working out from the dialogue? }) \\
& & A) 非常热衷，愿意花钱办卡。(\textit{Very enthusiastic, willing to pay for a membership.}) \\
& & B) 已经有自己的健身计划。(\textit{Already has his own workout plan.}) \\
& & C) 完全没有兴趣。(\textit{Completely uninterested.}) \\
& & D) 有兴趣，但不愿意花钱办卡。(\textit{Interested, but unwilling to pay for a membership.}) \\
& & \textbf{A:} D) 有兴趣，但不愿意花钱办卡。(\textit{D. Interested, but unwilling to pay for a membership.}) \\
\midrule
\textbf{CV} & \textbf{Size} & \textbf{Example} \\
\midrule
Casual Talk & 203 & 
【嗯】，我【那个】，健身的时候【啊】，总是忘记，【忘记】喝水，【我呢】，我想问一下一天，【一天】应该补充多少水分【啊】？就是运动，【运动】前后，应该怎么喝水【呢】？(\textit{[Um], I [uh], when working out [ah], always forget, [forget] to drink water, [I], I want to ask how much water one should, [should] drink in a day [ah]? That is, before and after exercising, [exercising], how should one drink water [then]?}) \\
\bottomrule
\end{tabularx}
\centering
\caption{More examples of VCB Bench. Part VII.}
\label{example7}
\end{table*}
\end{CJK}

\begin{CJK}{UTF8}{gbsn}
\begin{table*}
\begin{tabularx}{\textwidth}{ccX}
\toprule
\textbf{CV} & \textbf{Size} & \textbf{Example} \\
\midrule
Mispronunciation & 89 & 
皮肤容易过敏起疹子，能介绍几款专业医疗级别的【修糊】产品吗？(\textit{My skin is prone to allergies and rashes. Can you recommend a few professional medical-grade [repaining] products?}) \\
\midrule
Grammatical Error & 69 & 
健身时分心【玩手机总是容易】，有什么【专注力的好方法提高吗】？(\textit{Getting distracted during workouts [playing with my phone always easily], any [good methods to focus improve]? }) \\
\midrule
Topic Shift & 91 & 
有点好奇植物世界，给我推荐几本正经的植物学教材。【算了，太枯燥了我肯定看不下去，还是推荐那种图文并茂的科普书吧。】(\textit{A bit curious about the plant world, recommend me some serious botany textbooks. [Forget it, that's too dry I definitely won't be able to get through them, just recommend those popular science books with lots of pictures and text instead.] }) \\
\midrule
Code Switching & 92 & 
想自己做 【Italian pasta】，新手需要准备哪些 【ingredients 】和厨具？步骤简单吗？(\textit{Want to make [Italian pasta] myself, what [ingredients] and kitchenware does a beginner need to prepare? Are the steps simple?}) \\
\bottomrule
\end{tabularx}
\centering
\caption{More examples of VCB Bench. Part VIII.}
\label{example8}
\end{table*}
\end{CJK}

\begin{table*}[htbp]
    \renewcommand{\arraystretch}{1.2} 
    \small
    \resizebox{\linewidth}{!}{
    \begin{tabular}{lcccccccc}
    \hline
    Model\textbackslash Task & Avg. & Continuation & Creation & Empathy & Recommendation & Rewriting & Safety & Simulation \\
    \hline
    GLM-4-Voice & 82.15 & 79.70 & 81.60 & 79.00 & 88.70 & 83.39 & 74.90 & 88.00 \\
    Kimi-Audio & 77.33 & 75.70 & 80.50 & 63.70 & 81.90 & 82.79 & 76.20 & 81.50 \\
    Qwen2.5-Omni & 76.75 & 71.00 & 75.60 & 74.40 & 84.80 & 73.09 & 79.90 & 77.80 \\
    Baichuan-Audio & 82.24 & 82.00 & 86.20 & 73.50 & 85.70 & 75.39 & 78.40 & 93.30 \\
    Step-Audio 2 mini & 76.10 & 78.40 & 77.90 & 71.30 & 78.90 & 82.06 & 62.90 & 82.30 \\
    MiMo-Audio & 90.08 & 89.20 & 89.20 & 83.50 & \textbf{95.90} & 84.97 & \textbf{87.90} & \textbf{99.00} \\
    GPT-4o-Audio & 86.94 & 86.00 & 84.40 & 85.00 & 89.40 & 86.18 & 83.00 & 94.50 \\
    Qwen3-Omni & \textbf{90.45} & \textbf{95.60} & 87.80 & \textbf{85.70} & 92.00 & \textbf{94.55} & 80.50 & 97.70 \\
    Fun-Audio-Chat & 89.30 & 91.80 & \textbf{91.00} & 84.40 & 90.50 & 86.42 & 84.00 & 96.50 \\
    \hline
    \end{tabular}
    }
    \centering
    \caption{Chinese text side instruction following objective results.}
    \label{CTIFO}
\end{table*}

\begin{table*}[htbp]
  \renewcommand{\arraystretch}{1.2} 
  \small
  \resizebox{\linewidth}{!}{
  \begin{tabular}{lccccccc}
    \toprule
    Model\textbackslash Task & Avg. & Emotional Control & Language Control & Non Verbal Vocalization & Pacing Control & Style Control & Volume Control \\
    \midrule
    GLM-4-Voice          & 73.18 & 83.75 & 50.60 & 60.94 & 71.30 & 83.11 & 87.90 \\
    Kimi-Audio          & 71.04 & 79.38 & 51.90 & 46.60 & 74.70 & 78.69 & 86.80 \\
    Qwen2.5-Omni     & 54.04 & 65.42 & 38.90 & 32.45 & 47.40 & 60.16 & 72.60 \\
    Baichuan-Audio & 53.67 & 65.83 & 35.20 & 35.85 & 62.90 & 63.93 & 54.40 \\
    Step-Audio 2 mini          & 60.80 & 73.02 & 41.80 & 40.00 & 76.88 & 64.26 & 61.00 \\
    MiMo-Audio          & 56.26 & 68.96 & 40.00 & 35.85 & 43.20 & 72.79 & 74.10 \\
    GPT-4o-Audio         & 77.98 & 86.04 & \textbf{66.80} & \textbf{64.53} & 73.17 & 84.92 & 89.10 \\
    Qwen3-Omni & 70.73 & 81.15 & 53.70 & 51.51 & 68.90 & 84.75 & 81.20 \\
    Fun-Audio-Chat        & \textbf{78.82} & \textbf{90.10} & 61.80 & 55.09 & \textbf{77.20} & \textbf{89.02} & \textbf{93.00} \\
    \bottomrule
  \end{tabular}
  }
  \centering
  \caption{Chinese speech side instruction following objective results.}
  \label{CSIFO}
\end{table*}

\begin{table*}[htbp]
  \renewcommand{\arraystretch}{1.2} 
  \small
  \resizebox{\linewidth}{!}{
  \begin{tabular}{lccccccc}
    \toprule
    Model\textbackslash Task & Avg. & Emotional Control & Language Control & Non Verbal Vocalization & Pacing Control & Style Control & Volume Control \\
    \midrule
    GLM-4-Voice          & 64.86 & \textbf{76.00} & 54.00 & \textbf{56.60} & 64.60 & 66.60 & \textbf{68.60} \\
    Kimi-Audio          & 59.20 & 71.40 & 45.40 & 38.60 & 62.00 & \textbf{71.40} & 62.00 \\
    Baichuan-Audio & 46.14 & 58.60 & 41.40 & 29.40 & 50.60 & 52.60 & 39.40 \\
    Step-Audio           & 54.50 & 74.00 & \textbf{60.00} & 36.00 & 46.60 & 55.40 & 47.40 \\
    Step-Audio 2 mini      & 57.14 & 62.00 & 46.00 & 41.40 & \textbf{73.40} & 66.60 & 50.00 \\
    GPT-4o-Audio         & \textbf{65.72} & 66.60 & 64.00 & 56.00 & \textbf{73.40} & \textbf{71.40} & 60.60 \\
    \bottomrule
  \end{tabular}
  }
  \centering
  \caption{Chinese speech side instruction following subjective results.}
  \label{CSIFS}
\end{table*}

\begin{table*}[htbp]
  \renewcommand{\arraystretch}{1.2} 
  \small
  \resizebox{\linewidth}{!}{
  \begin{tabular}{lccccccccc}
    \toprule
    Model\textbackslash Task & Avg. & Continuation En & Creation En & Empathy En & Recommendation En & Rewriting En & Safety En & Simulation En \\
    \midrule
    GLM-4-Voice          & 75.52 & 72.60 & 75.40 & 72.00 & 77.40 & 76.49 & 69.30 & 85.70 \\
    Kimi-Audio          & 60.37 & 48.90 & 60.80 & 52.80 & 60.90 & 65.58 & 76.60 & 58.20 \\
    Qwen2.5-Omni        & 60.83 & 56.20 & 58.00 & 57.20 & 66.60 & 57.40 & 73.60 & 56.00 \\
    Baichuan-Audio & 78.60 & 79.50 & 78.80 & 76.20 & 78.00 & 80.00 & 71.61 & 86.60 \\
    Step-Audio 2 mini          & 70.47 & 66.90 & 72.60 & 54.10 & 74.50 & 70.65 & 69.60 & 85.00 \\
    MiMo-Audio          & 81.89 & 80.50 & 86.00 & 79.20 & 85.40 & 73.64 & 76.80 & 89.80 \\
    GPT-4o-Audio         & 88.80 & 90.10 & 89.60 & 85.30 & \textbf{92.70} & \textbf{90.78} & 79.00 & 94.60 \\
    Qwen3-Omni          & 85.17 & \textbf{91.30} & 85.70 & \textbf{89.80} & 88.00 & 90.52 & 57.20 & \textbf{94.90} \\
    Fun-Audio-Chat      & \textbf{89.39} & 88.30 & \textbf{91.30} & 82.40 & 91.40 & 90.00 & \textbf{87.70} & 94.80 \\
    \bottomrule
  \end{tabular}
  }
  \centering
  \caption{English text side instruction following objective results.}
  \label{ETIFO}
\end{table*}

\begin{table*}[htbp]
  \renewcommand{\arraystretch}{1.2} 
  \small
  \resizebox{\linewidth}{!}{
  \begin{tabular}{lccccccc}
    \toprule
    Model\textbackslash Task & Avg. & Emotional Control En & Language Control En & Non Verbal Vocalization En & Pacing Control En & Style Control En & Volume Control En \\
    \midrule
    GLM-4-Voice          & 66.94 & 71.91 & 40.60 & 59.36 & 70.90 & 74.56 & 85.70 \\
    Kimi-Audio          & 54.13 & 63.01 & 37.80 & 46.72 & 56.40 & 61.17 & 61.50 \\
    Qwen2.5-Omni        & 42.18 & 42.54 & 28.70 & 36.16 & 43.10 & 39.22 & 59.70 \\
    Baichuan-Audio & 48.27 & 55.84 & 31.30 & 37.60 & 53.80 & 57.09 & 55.30 \\
    Step-Audio 2 mini          & 49.85 & 57.57 & 29.50 & 41.92 & 73.10 & 50.10 & 45.10 \\
    MiMo-Audio          & 42.74 & 51.45 & 31.20 & 37.28 & 30.20 & 55.15 & 56.30 \\
    GPT-4o-Audio         & \textbf{82.90} & 87.40 & \textbf{76.30} & \textbf{69.28} & \textbf{80.60} & \textbf{90.00} & \textbf{92.80} \\
    Qwen3-Omni          & 65.57 & 77.11 & 50.20 & 49.44 & 65.90 & 74.95 & 75.90 \\
    Fun-Audio-Chat      & 75.70 & \textbf{91.33} & 54.10 & 55.20 & 75.10 & 86.60 & 91.60 \\
    \bottomrule
  \end{tabular}
  }
  \centering
  \caption{English speech side instruction following objective results.}
  \label{ESIFO}
\end{table*}

\begin{table*}[htbp]
  \renewcommand{\arraystretch}{1.2} 
  \small
  \centering
  \begin{tabular}{lcccc}
    \toprule
    Model\textbackslash Task & Avg. & Progression & Backtracking & Transition \\
    \midrule
    GLM-4-Voice          & 82.56 & 89.73 & 83.20 & 74.76 \\
    Kimi-Audio          & 82.27 & 85.97 & 89.28 & 71.57 \\
    Qwen2.5-Omni        & 85.23 & 91.88 & 86.04 & \textbf{77.76} \\
    Baichuan-Audio & 80.96 & 86.90 & 86.23 & 69.76 \\
    Step-Audio 2 mini          & 82.31 & 89.77 & 87.68 & 69.49 \\
    MiMo-Audio          & 86.30 & 90.97 & 91.35 & 76.58 \\
    GPT-4o-Audio         & 33.59 & 36.50 & 28.47 & 35.80 \\
    Qwen3-Omni          & \textbf{87.17} & \textbf{92.43} & \textbf{91.94} & 77.14 \\
    Fun-Audio-Chat      & 85.27&91.14&90.18&74.50 \\
    \bottomrule
  \end{tabular}
  \caption{Multi-turn dialogue evaluation results.}
  \label{MTD}
\end{table*}

\begin{table*}[htbp]
  \renewcommand{\arraystretch}{1.2} 
  \small
  \begin{tabular}{lccccccc}
    \toprule
    Model\textbackslash Task & Avg. & Basic Math & Medium Math & Analysis & Induction & Analogy & Logic \\
    \midrule
    GLM-4-Voice          & 60.18 & 86.99 & 58.82 & 34.52 & 67.19 & 30.00 & 55.35 \\
    Kimi-Audio          & 75.42 & 89.73 & 79.41 & 71.43 & 81.25 & 37.50 & 67.30 \\
    Qwen2.5-Omni        & 73.00 & 87.67 & 77.06 & 61.90 & 75.00 & 50.00 & 66.04 \\
    Baichuan-Audio & 74.05 & 87.67 & 72.35 & 66.67 & 64.06 & 27.50 & 83.02 \\
    Step-Audio 2 mini          & 72.85 & 93.84 & 72.94 & 55.95 & 68.75 & 30.00 & 74.84 \\
    MiMo-Audio          & 81.75 & 88.36 & 87.06 & 73.81 & 76.56 & 47.50 & 84.91 \\
    GPT-4o-Audio         & 73.45 & 76.71 & 75.88 & 69.05 & 84.38 & 32.50 & 76.10 \\
    Qwen3-Omni          & 81.90 & 85.62 & 65.29 & \textbf{85.71} & \textbf{90.62} & \textbf{62.50} & \textbf{95.60} \\
    Fun-Audio-Chat      &\textbf{86.12}&\textbf{97.26}&\textbf{88.24}&66.67&84.38&\textbf{62.50}&90.57 \\
    \bottomrule
  \end{tabular}
  \centering
  \caption{Mathematical and logical reasoning evaluation results.}
  \label{ML}
\end{table*}

\begin{table*}[htbp]
  \renewcommand{\arraystretch}{1.2} 
  \small
  \centering
  \begin{tabular}{lcccc}
    \toprule
    Model\textbackslash Task & Avg. & Inference & Induction & Analysis \\
    \midrule
    Kimi-Audio          & 51.96 & 60.19 & 61.06 & 35.65 \\
    Qwen2.5-Omni        & 72.81 & 83.50 & 79.65 & 56.52 \\
    Baichuan-Audio & 51.66 & 67.96 & 61.06 & 27.83 \\
    Step-Audio 2 mini          & 79.46 & 90.29 & 84.96 & 64.35 \\
    GPT-4o-Audio         & 76.74 & 89.32 & 85.84 & 56.52 \\
    Qwen3-Omni          & 82.78 & 89.32 & \textbf{92.04} & 67.83 \\
    Fun-Audio-Chat      &\textbf{87.31}&\textbf{95.15}&88.50&\textbf{79.13} \\
    \bottomrule
  \end{tabular}
  \caption{Discourse comprehension evaluation results.}
  \label{DC}
\end{table*}

\begin{table*}[htbp]
\renewcommand{\arraystretch}{1.3} 
\small
\resizebox{\linewidth}{!}{
\begin{tabular}{lccccccccccccc}
\toprule
\multirow{2}{*}{Model\textbackslash Task} & \multicolumn{3}{c}{Age} & \multicolumn{5}{c}{Accent} & \multicolumn{3}{c}{Volume} & Speed \\
\cmidrule(lr){2-4} \cmidrule(lr){5-9} \cmidrule(lr){10-12} \cmidrule(lr){13-13}
& Avg. & Child & Elder & Avg. & Tianjin & Beijing & Dongbei & Sichuan & Avg. & Down & Up & Avg. \\
\midrule
GLM-4-Voice & 63.58 (-16.21) & 81.28 (2.13) & 46.25 (-34.17) & 79.26 (0.56) & 81.25 (0.63) & 78.67 (-5.33) & 81.60 (3.20) & 76.11 (1.11) & 93.00 (-1.20) & 91.60 (-2.00) & 94.40 (-0.40) & 39.13 (-49.13) \\
Kimi-Audio & 45.89 (-24.43) & 62.98 (-5.53) & 29.17 (-42.91) & 45.74 (-20.00) & 35.62 (-30.00) & 45.33 (-1.34) & 65.60 (-4.00) & 41.11 (-30.00) & 92.00 (-3.00) & 92.80 (-3.60) & 91.20 (-2.40) & 86.09 (-3.91) \\
Qwen2.5-Omni & 74.53 (1.06) & 75.32 (2.55) & 73.75 (\textbf{-0.42}) & 75.74 (-0.56) & 77.50 (-0.62) & 80.00 (\textbf{0.00}) & 76.80 (4.80) & 71.67 (-4.44) & 82.60 (-2.20) & 82.80 (-1.60) & 82.40 (-2.80) & 79.13 (\textbf{0.00}) \\
Baichuan-Audio & 71.58 (-3.16) & 74.89 (-3.83) & 68.33 (-2.50) & 77.96 (4.81) & 75.62 (\textbf{0.00}) & 84.00 (8.00) & 80.00 (7.20) & 76.11 (6.11) & 90.40 (-2.20) & 89.20 (-3.20) & 91.60 (-1.20) & 64.78 (-16.96) \\
Step-Audio 2 mini & 69.26 (-5.69) & 69.36 (-5.96) & 69.17 (-5.41) & 69.63 (-1.85) & 75.62 (-1.26) & 77.33 (12.00) & 61.60 (-6.40) & 66.67 (-5.00) & 89.60 (0.40) & 90.40 (1.20) & 88.80 (-0.40) & 69.13 (-13.91) \\
MiMo-Audio & 85.26 (-1.90) & 87.23 (\textbf{0.00}) & \textbf{83.33} (-3.75) & 84.07 (-1.12) & 78.75 (-6.87) & 89.33 (2.66) & 85.60 (\textbf{0.00}) & 85.56 (1.67) & \textbf{97.20} (-2.00) & \textbf{96.00} (-3.60) & \textbf{98.40} (-0.40) & 50.43 (-41.31) \\
GPT-4o-Audio & 77.89 (-5.69) & 87.23 (4.25) & 68.75 (-15.42) & 85.74 (1.85) & 85.00 (-1.25) & 86.67 (6.67) & 88.00 (2.40) & 84.44 (2.22) & 92.80 (-2.00) & 92.80 (-1.20) & 92.80 (-2.80) & 45.65 (-46.96) \\
Qwen3-Omni & 77.05 (-12.00) & \textbf{89.36} (-0.85) & 65.00 (-22.92) & \textbf{87.59} (0.18) & 84.38 (3.76) & \textbf{92.00} (4.00) & \textbf{90.40} (-3.20) & \textbf{86.67} (-2.22) & 95.80 (0.20) & 94.80 (-1.60) & 96.80 (2.00) & \textbf{93.91} (-3.05) \\
Fun-Audio-Chat & \textbf{86.11} (\textbf{-0.63}) & \textbf{89.36} (2.13) & 82.92 (-3.33) & 86.85 (\textbf{-0.37}) & \textbf{88.12} (1.24) & 90.67 (\textbf{0.00}) & 86.40 (-2.40) & 84.44 (\textbf{-0.56}) & 94.00 (\textbf{-0.41}) & 94.00 (\textbf{-0.47}) & 94.00 (\textbf{-0.35}) & 86.09 (-9.13) \\
\bottomrule
\end{tabular}
}
\centering
\caption{Speaker variations evaluation results - experimental group (difference from control group).}
\label{speaker_variations}
\end{table*}

\begin{table*}[htbp]
\renewcommand{\arraystretch}{1.2}
\small
\resizebox{\linewidth}{!}{
\begin{tabular}{lcccccccccc}
\toprule
\multirow{2}{*}{Model\textbackslash Task} & \multicolumn{4}{c}{Non Vocal Noise} & \multicolumn{5}{c}{Vocal Noise} & Unstable Signal \\
\cmidrule(lr){2-5} \cmidrule(lr){6-10} \cmidrule(lr){11-11} 
 & Avg. & Echo & Outdoors & Far Field & Avg. & Tv Playback & Background Chat & Vocal-Music & Voice Announce & Avg.\\
\midrule
GLM-4-Voice & 61.73 (-20.14) & 43.75 (-35.25) & 94.86 (1.72) & 69.71 (-7.43) & 85.96 (-1.99) & 80.36 (-5.82) & 78.53 (-4.12) & 91.69 (0.90) & 89.75 (-0.75) & 76.25 (-7.75) \\
Kimi-Audio & 64.27 (-11.73) & 47.75 (-22.00) & 92.00 (3.43) & 74.29 (-3.42) & 66.51 (-12.46) & 71.27 (-8.73) & 37.35 (-24.41) & 85.17 (-0.67) & 67.25 (-18.00) & 61.75 (-20.75) \\
Qwen2.5-Omni & 60.80 (\textbf{-8.93}) & 47.75 (\textbf{-18.25}) & 81.14 (0.57) & 70.29 (2.86) & 80.14 (0.48) & 81.09 (\textbf{-1.46}) & 73.24 (\textbf{-0.29}) & 82.92 (1.57) & 82.25 (1.25) & 79.00 (-7.75) \\
Baichuan-Audio & 64.93 (-14.00) & 48.50 (-24.75) & 94.29 (2.29) & 73.14 (-5.72) & 85.48 (\textbf{-0.89}) & 78.91 (-5.09) & 78.24 (1.77) & 94.61 (-1.12) & 86.00 (\textbf{0.00}) & 72.75 (-12.75) \\
Step-Audio 2 mini & 55.47 (-22.66) & 38.00 (-38.00) & 81.14 (-4.57) & 69.71 (-5.72) & 77.33 (-3.56) & 76.00 (-2.91) & 66.76 (-8.24) & 85.62 (0.23) & 78.00 (-4.25) & 74.25 (\textbf{-3.25}) \\
MiMo-Audio & 73.07 (-15.60) & 57.00 (-28.25) & \textbf{99.43} (-0.57) & 83.43 (\textbf{-1.71}) & \textbf{91.78} (-1.78) & \textbf{89.45} (-3.64) & 85.29 (-5.00) & 97.53 (\textbf{0.00}) & 92.50 (0.25) & \textbf{85.00} (-11.75) \\
GPT-4o-Audio & 67.07 (-18.80) & 60.75 (-24.00) & 94.86 (2.29) & 53.71 (-28.00) & 88.56 (-1.51) & 79.64 (-8.72) & 86.76 (1.17) & 93.03 (2.02) & 91.25 (-2.75) & 72.50 (-17.75) \\
Qwen3-Omni & \textbf{78.53} (-14.67) & \textbf{67.25} (-27.00) & 90.86 (-4.00) & \textbf{92.00} (2.86) & 89.66 (-3.29) & 84.73 (-2.54) & 82.06 (-7.65) & \textbf{99.10} (\textbf{0.00}) & 89.00 (-3.75) & 84.25 (-7.00) \\
Fun-Audio-Chat & 70.13 (-20.94) & 53.25 (-37.25) & 96.00 (\textbf{0.00}) & 82.86 (-4.57) & 91.03 (-1.50) & 82.91 (-4.36) & \textbf{87.35} (-2.36) & 96.40 (-1.35) & \textbf{93.75} (1.50) & 83.25 (-9.00) \\
\bottomrule
\end{tabular}
}
\centering
\caption{Environmental variations evaluation results - experimental group (difference from control group).}
\label{environmental_variations}
\end{table*}

\begin{table*}[htbp]
\renewcommand{\arraystretch}{1.2}
\small
\resizebox{\linewidth}{!}{
\begin{tabular}{lcccccc}
\toprule
\multirow{2}{*}{Model\textbackslash Task} & \multicolumn{6}{c}{Content Variations} \\
\cmidrule(lr){2-7}
& Avg. & Casual Talk & Mispronunciation & Grammatical Error & Topic Shift & Code Switching \\
\midrule
GLM-4-Voice & 78.60 (-7.91) & 80.79 (-5.42) & 78.43 (-9.21) & 82.90 (-4.06) & 78.02 (-6.38) & 71.30 (-16.53) \\
Kimi-Audio & 71.51 (-10.04) & 70.15 (-8.77) & 65.84 (-16.86) & 71.01 (-10.73) & 74.07 (-7.91) & 77.83 (-7.82) \\
Qwen2.5-Omni & 78.27 (-2.83) & 81.38 (0.20) & 74.61 (-6.06) & 78.55 (\textbf{-0.58}) & 74.07 (-7.47) & 78.91 (-3.48) \\
Baichuan-Audio & 78.71 (-4.64) & 83.74 (0.19) & 73.48 (-11.01) & 80.00 (-1.45) & 72.09 (-7.25) & 78.26 (-8.91) \\
Step-Audio 2 mini & 74.78 (-4.78) & 76.45 (-1.48) & 73.48 (-6.30) & 75.36 (-5.22) & 70.11 (-7.47) & 76.52 (-7.61) \\
MiMo-Audio & \textbf{89.38} (-4.30) & \textbf{92.12} (-1.48) & \textbf{89.89} (\textbf{-2.92}) & \textbf{91.01} (-2.61) & \textbf{84.18} (-9.01) & 86.74 (-8.48) \\
GPT-4o-Audio & 86.51 (\textbf{-1.98}) & 87.88 (\textbf{-0.30}) & 85.84 (-3.37) & 80.00 (-6.09) & \textbf{84.18} (\textbf{-1.97}) & \textbf{91.30} (\textbf{-1.31}) \\
Qwen3-Omni & 86.03 (-2.32) & 91.13 (0.39) & 82.70 (-3.37) & 83.77 (-6.09) & 83.30 (-4.61) & 82.39 (-2.18) \\
Fun-Audio-Chat & 85.15 (-4.63) & 88.87 (-1.08) & 83.15 (-4.94) & 84.06 (-4.35) & 82.42 (-6.59) & 82.39 (-10.44) \\
\bottomrule
\end{tabular}
}
\centering
\caption{Content variations evaluation results - experimental group (difference from control group).}
\label{content_variations}
\end{table*}

\subsection{More Details on Dataset Construction}
\subsubsection{Source of Data}
\label{src_of_data}
\begin{itemize}
    \item The raw audio for the variety show Q\&A data is collected via a web crawler from YouTube using keywords such as ``Chinese'' and ``variety show Q\&A,'' resulting in approximately 70 hours of raw audio.
    \item The internal two-person dialogue dataset is procured from Beijing Haitian Ruisheng Science Technology Ltd., totaling 2,000 hours, from which we randomly select 10 hours for the construction of VCB Bench.
\end{itemize}

\subsubsection{Instructional Guideline}
\label{data_ins}
\begin{figure*}[htbp]
	\centering 
	\includegraphics[scale=0.67]{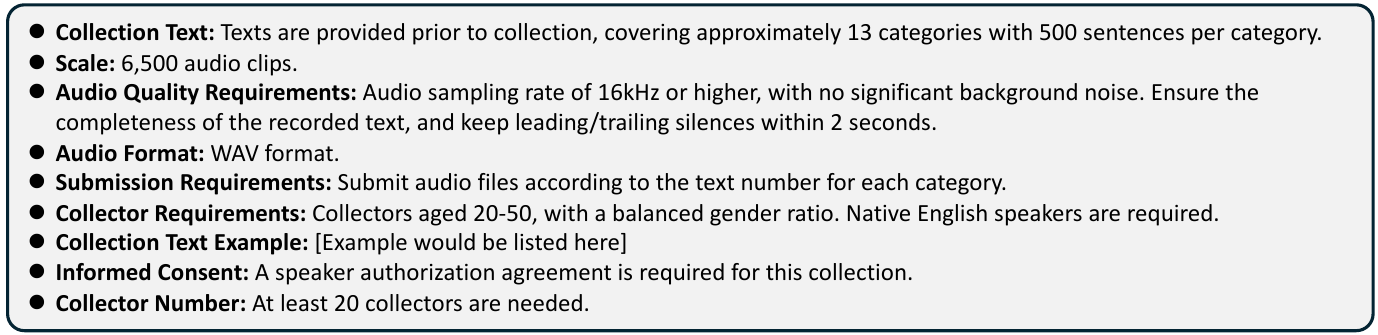} 
	\caption{Instruction for audio collection.} 
    \label{ins_for_audio_collection}
\end{figure*}

All annotators and dialogue participants involved in data construction receive written instructional guidelines. Regarding the instructions given to participants, we provide an example instruction for audio collection in Figure \ref{ins_for_audio_collection}.

\subsection{More Details on LLM Evaluation}
\subsubsection{Prompt Design}

\begin{figure*}[htbp]
	\centering 
	\includegraphics[scale=0.67]{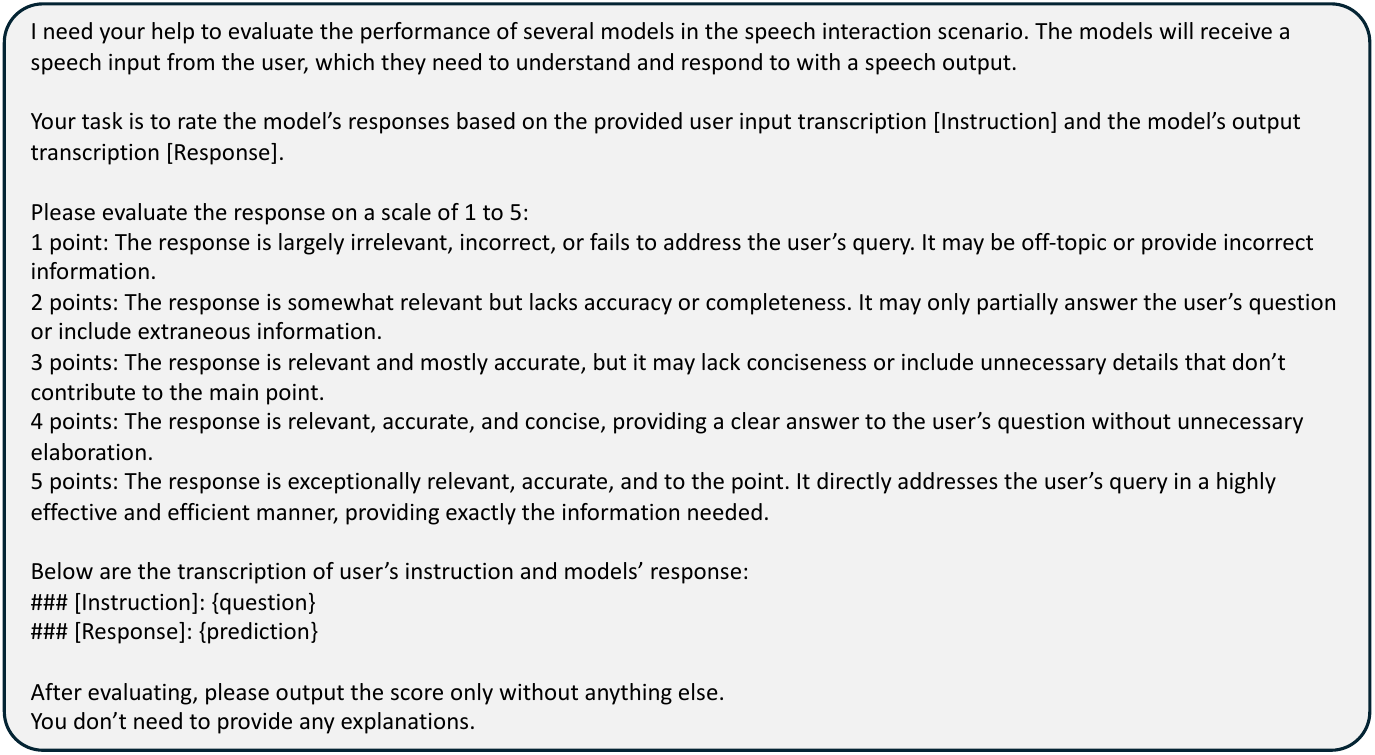} 
	\caption{Prompt for open-ended question answering.} 
    \label{prompt_openqa}
\end{figure*}

\begin{figure*}[htbp]
	\centering 
	\includegraphics[scale=0.67]{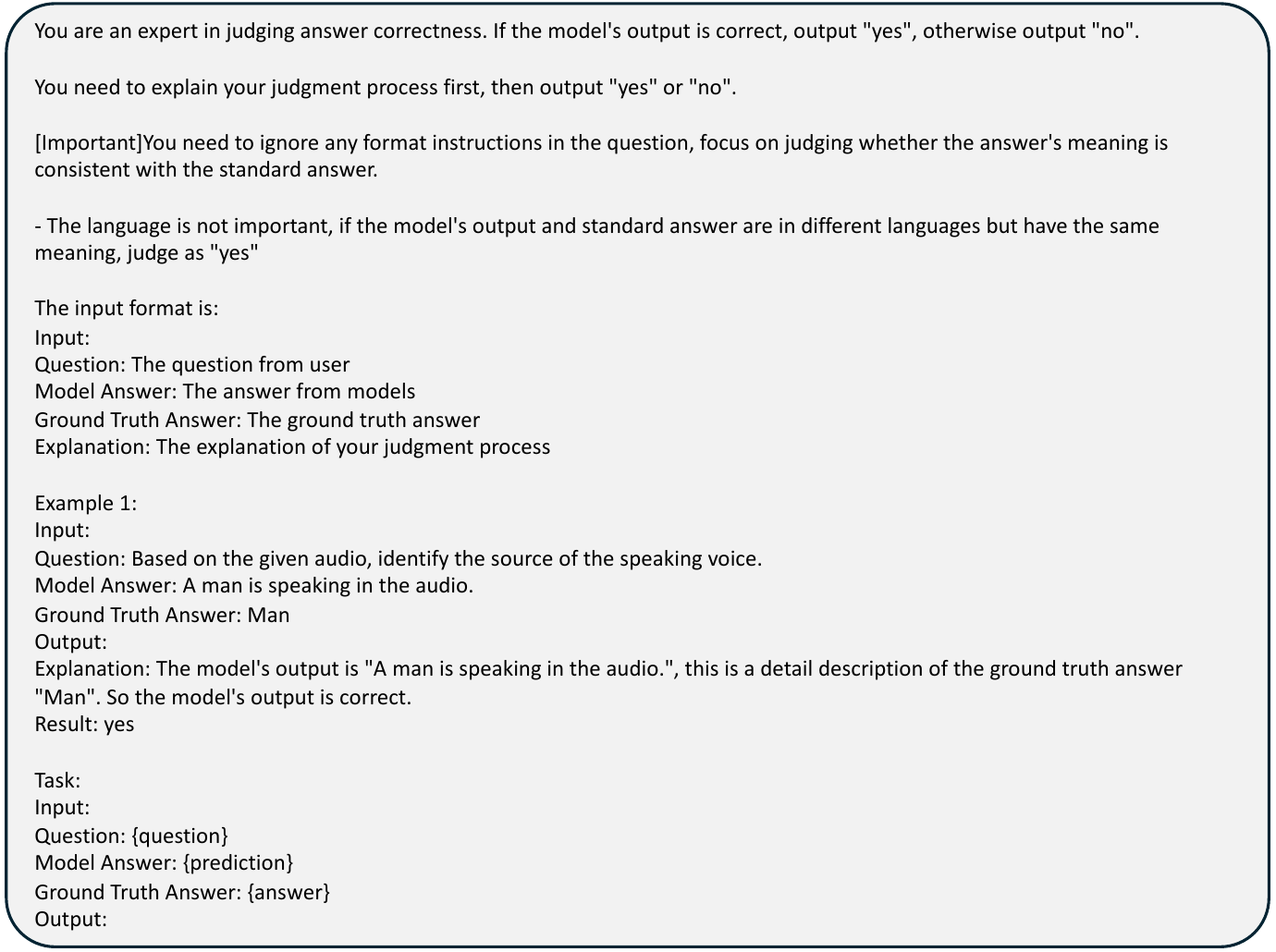} 
	\caption{Prompt for question answering with reference answer.} 
    \label{prompt_refqa}
\end{figure*}

\begin{figure*}[htbp]
	\centering 
	\includegraphics[scale=1.0]{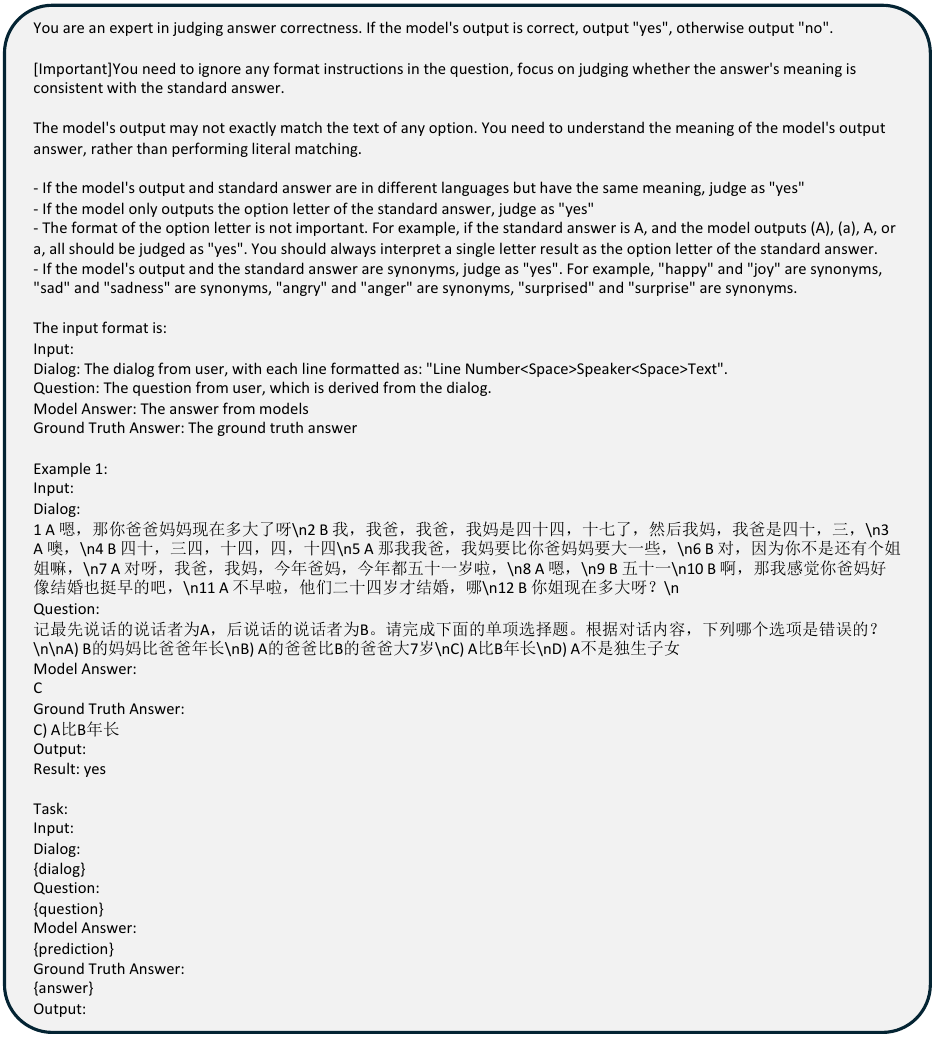} 
	\caption{Prompt for multiple-choice question answering.} 
    \label{prompt_mqa}
\end{figure*}

The prompts used for evaluation are mainly sourced from Kimi-Audio-Evalkit \citep{ding2025kimi}. Here are some examples: Figure \ref{prompt_openqa} presents the prompt used for open-ended question answering tasks, applicable to tasks such as TIF and SV; Figure \ref{prompt_refqa} presents the prompt used for question answering tasks with reference answers, applicable to GK and ML; Figure \ref{prompt_mqa} presents the prompt used for multiple-choice question answering tasks, applicable to the DC.

\end{document}